\documentclass[10pt,journal,compsoc]{IEEEtran}
\usepackage{amsmath,amssymb,graphicx}
\usepackage{url,subfigure,epstopdf}
\usepackage{graphics}
\usepackage{epsfig}
\usepackage[usenames,dvipsnames]{color}
\usepackage{comment}
\usepackage{cases}
\usepackage{bbm}

\ifCLASSOPTIONcompsoc
  \usepackage[nocompress]{cite}
\else
  \usepackage{cite}
\fi

\DeclareMathAlphabet\mathbfcal{OMS}{cmsy}{b}{n}

\DeclareMathOperator*{\argmax}{arg\,max}

\newcommand{\dd}{{\rm\,d}} 
\newcommand{\ee}{{\rm e}}

\newcommand{\1}{\mathbbm 1}

\newcommand{\Ac}{\mathcal A}
\newcommand{\Bc}{\mathcal B}

\newcommand{\Pc}{\mathcal P}

\newcommand{\Uc}{\mathcal U}

\newcommand{\Xc}{\mathcal X}
\newcommand{\Zc}{\mathcal Z}

\newtheorem{teorema}{\bf Theorem}

\newcommand{\plus}{\mathord{+}}
\newcommand{\minus}{\mathord{-}}

\newcommand{\RR}{\mathbb{R}}

\newcommand{\EE}{\mathbb{E}} 

\newcommand{\ls}[1]
   {\dimen0=\fontdimen6\the\font
    \lineskip=#1\dimen0
    \advance\lineskip.5\fontdimen5\the\font
    \advance\lineskip-\dimen0
    \lineskiplimit=.9\lineskip
    \baselineskip=\lineskip
    \advance\baselineskip\dimen0
    \normallineskip\lineskip
    \normallineskiplimit\lineskiplimit
    \normalbaselineskip\baselineskip
    \ignorespaces
}
\makeatother

\begin{document}

\title{Belief Dynamics in Social Networks:\\ A Fluid-Based Analysis}

\author{Alessandro~Nordio,~\IEEEmembership{Member,~IEEE,}
        Alberto~Tarable,~\IEEEmembership{Member,~IEEE,}
        Carla~Fabiana~Chiasserini,~\IEEEmembership{Senior~Member,~IEEE,}
        and~Emilio~Leonardi,~\IEEEmembership{Senior~Member,~IEEE}
\IEEEcompsocitemizethanks{\IEEEcompsocthanksitem A. Nordio is with the Institute of Electronics, Computer and Telecommunication Engineering, National Research Council (CNR-IEIIT), Torino, Italy.\protect\\
E-mail: alessandro.nordio@ieiit.cnr.it
\IEEEcompsocthanksitem A. Tarable is with the Institute of Electronics, Computer and Telecommunication Engineering, National Research Council (CNR-IEIIT), Torino, Italy.\protect\\
E-mail: alberto.tarable@ieiit.cnr.it
\IEEEcompsocthanksitem C. F. Chiasserini is with the Department of Electronics and Telecommunications (DET), Politecnico di Torino and with CNR-IEIIT, Torino, Italy.\protect\\
E-mail: carla.chiasserini@polito.it
\IEEEcompsocthanksitem E. Leonardi is with the Department of Electronics and Telecommunications (DET), Politecnico di Torino and with CNR-IEIIT, Torino, Italy.\protect\\
E-mail: emilio.leonardi@polito.it
}
}

\IEEEtitleabstractindextext{%
\begin{abstract}
The advent and proliferation of social media have led to the development of
mathematical models describing the evolution of beliefs/opinions in an ecosystem
composed of socially interacting users. The goal is to gain insights into
collective dominant social beliefs and into the impact of different
components of the system, such as users' interactions, while being able
to predict users' opinions.  Following this thread, in this paper we
consider a fairly general dynamical model of social interactions, which captures
all the main features exhibited by a social system.  For such  model, by
embracing a mean-field approach, we derive a diffusion differential
equation that represents asymptotic belief dynamics, as the number of
users grows large.  We then analyze the steady-state behavior as well as
the time dependent (transient) behavior of the system.  In particular, for the
steady-state distribution, we obtain simple closed-form expressions for a
relevant class of systems, while we propose efficient semi-analytical techniques
in the most general cases.  At last, we develop an efficient semi-analytical
method to analyze the dynamics of the users' belief over time, which can be
applied to a remarkably large class of systems.
\end{abstract}}

\maketitle
\pagestyle{plain}

\section{Introduction\label{sec:intro}}

Since the advent and proliferation of social media, the
  research community has devoted significant effort to develop
  mathematical models describing the evolution of beliefs/opinions in
  an ecosystem composed of socially interacting
  users~\cite{Colbaugh2010,Asur:2010,shi2013agreement,Baccelli}.  In
  addition, many enterprises and government agencies have shown great
  interest in using social media data with the aim to improve customer
  relationship as well as public opinion management.  For example,
understanding the sentiment in the public opinion allows an effective
management of the public response to natural disasters by clarifying
facts; political parties can use social media to sense people's
opinion about their actions \cite{Scaglione,NIPS2014}; knowledge of
brand sentiment acquired through social sites can lead to effective
marketing campaigns.

In this context, several approaches 
to opinion sensing, based on sentiment analysis \cite{Pang2008}, have
been recently presented. Furthermore, 
several studies,
e.g.,~\cite{degroot1974reaching,F&J,Fagnani-Acemoglu,shi2013agreement,Baccelli,Garnier,Tempo}, 
have addressed the need to understand and forecast
belief dynamics by developing theoretical models. These models have provided important
insights into the impact of social interactions, as well as possible explanatory
mechanisms to the emergence of strong collective opinions. Interestingly, they
have also analyzed the impact of possible strategies to influence social
beliefs.

A typical way to model social interactions between users (hereinafter also
called agents) is to use graphs, either static or dynamic,
which reflect the social structure of the
system and how users interact.
In this  representation, often users directly interact only with
their neighbors, varying their beliefs for effect of pairwise ``attractive''
interactions~\cite{F&J,Tempo}.
In the case of social graph whose structure varies dynamically~\cite{shi2013agreement}
a class of models that have attained considerable popularity, is
represented by the so-called {\em bounded confidence}, in which interactions
between agents are effective only if the agents' beliefs are
sufficiently similar (i.e., the difference between their beliefs is below a
given prefixed threshold) \cite{7-Deffuant,8-Hegselmann}. On the one hand,
bounded confidence models are particularly interesting because they permit to
represent belief-dependent social behaviors, such as  \lq\lq homophily'',
which are often observed in real systems.  On the other hand, their analysis
poses several challenges because the equations driving the agents' interactions
become non linear 
\cite{Fagnani-Como,Garnier,9-Blondel,16-Weisbuch,17-Bhattacharyya,Varshney2014,Varshney2017}.

In this paper, we focus on developing a convenient and comprehensive model
of social belief dynamics which can account for bounded confidence. With respect to previous work  (discussed in detail in Section \ref{sec:rel-work}), we make a significant step forward. 
\begin{itemize}

\item[{\em (i)}] 
We generalize the model proposed in \cite{Baccelli}, which combines
 features such as constrained social interactions,  bounded confidence and agent endogenous opinion dynamics, by
introducing the agent's
prejudice. This is an important component  originally introduced by
Friedkin and Johnsen~\cite{F&J}, but 
neglected  in \cite{Baccelli}. We show
that the introduction of agents' prejudices ensures system stability (i.e.,
beliefs cannot drift to infinite) -- an important, amenable property for a
model of belief dynamics.

\item[{\em (ii)}]
  In order to represent the dynamics of the agents' belief over time,
  we develop an efficient method based on mean-field analysis, which
  applies to the case of a large number of agents and in absence of
  bounded confidence.  We also show how simple closed-form expressions
  for the steady-state distribution can be derived in this case.

\item[{\em (iii)}]
  In the general scenario where bounded confidence is in place, we
  give insights into the beliefs steady-state distribution, and, under
  mild assumptions, we provide a computationally efficient method to
  derive it.

\item[{\em (iv)}]
  We exploit our analytical results to show interesting belief
  dynamics in scenarios where agents exhibit different personalities
  and degree of stubbornness. In particular, we show
  the beliefs' temporal evolution right after a breaking news has been
  posted, and how the interaction between two different user
  communities affects opinions. The observed behaviors match those
  described by sociology studies such as~\cite{sociology-dramatic,
    sociology-community}.
\end{itemize}
 
\section{System model and properties\label{sec:model}}
We start by casting the beliefs' temporal evolution in a system
including a discrete set of agents, each of which may have a different
belief and personality. Then we let the number of agents grow large
and we define a continuous belief-personality bi-dimensional
space. Through such an asymptotic representation of the system, and by
by using a mean-field approach we derive the equation representing how
the probability density of agents varies over time in the
belief-personality space.

\subsection{Temporal evolution of agents' beliefs}
Consider a set of agents $\Uc$, with cardinality $U$, with agent $i$
exhibiting personality $P_i\in \Pc$. The agent's personality accounts
for the interests and the habits of a user, e.g., the social networks
to which she has subscribed or the forums in which she participates.
Agent $i\in \Uc$ has a belief $X_i(t) \in \Xc$, which evolves over
continuous time, $t\in \RR_+$.
We define the prejudice $u(P_i)$ as the a-priori belief of agent $i$,
which depends on the agent's personality.  The opportunity that agents
have to interact with each other is modeled through a graph
representing the existence and the intensity of social relationships
between users, which depend on the personality of the agents and on
the similarity between their beliefs.  The actual influence that
agents exert on each other then depends on the opportunity they have
to interact, as well as on their willingness to exchange beliefs.

As a result, the evolution of agent $i$'s belief over time can be
represented as:
\begin{eqnarray}
\label{eq:x_i}
X_i(t\plus\dd t)
&\mathord{=}& X_i(t)  \nonumber \\ 
& &\hspace{-13ex}+\frac{1\minus\alpha(P_i)}{U}\sum_{\substack{j\in \Uc \\ j \neq i}}  \zeta\left(|X_j(t)\minus X_i(t)|\mathord{,}P_i\mathord{,}P_j\right)\left[X_j(t)\minus X_i(t)\right] \dd t  \nonumber \\ 
&&\hspace{-13ex}+\alpha(P_i)\left[u(P_i)\mathord{-}X_i(t)\right] \dd t \nonumber \\
&&\hspace{-13ex}   + \sigma \dd W_i(t) \,.
\end{eqnarray}
The meaning of the terms in the right hand side (RHS) of the above
expression is as follows. 
\begin{itemize}

\item The first term denotes the belief of user $i$ at the current 
  time instant.

\item The second term  represents the interaction of agent $i$ with
  all other agents in $\Uc$. In particular,
  \begin{itemize}
    \item $\alpha(P_i)\in [0,1]$ indicates how sensitive $i$ is to
      other agents' beliefs, which, as also discussed in
      \cite{sociology-community}, plays an important role in opinion
      dynamics. This parameter will also be referred to as user's
      level of stubbornness. When $\alpha(P_i)\to 1$, the agent becomes
      completely insensitive to other beliefs (stubborn). Instead, as
      $\alpha(P_i)$ decreases, the agent is more inclined to
      accept others' beliefs and is less conditioned by her own
      prejudice. For brevity, in the following we denote
      $\bar{\alpha}(p) =1-\alpha(p)$;
    \item $\zeta(|X_j(t)\mathord{-}X_i(t)|,P_i,P_j)\ge 0$ represents
      the presence and the strength of interactions between agents $i$
      and $j$ (hereinafter also referred to as mutual influence).  In
      the most general case, it is a function of both agents'
      personality and of the
      distance between the agents' beliefs which define the structure of the social graph~\cite{sociology-community}. Note that, whenever
      $\zeta(\cdot,P_i,P_j)=0$, the two agents do not influence each
      other, i.e., two agents never interact. Also, it is fair to
      assume that $\zeta(|X_j(t)\mathord{-}X_i(t)|,P_i,P_j)$ is (i)
      upper bounded by a constant and (ii) a smooth function (i.e., it
      has at least the first derivative continuous everywhere) with
      respect to its first argument.
   \end{itemize}

\item The third term represents the tendency of an agent to retain her
  prejudice.

\item The fourth term accounts for the endogenous process of the belief
  evolution within each user. Such process is modeled as an
  i.i.d. standard Brownian motion with zero drift and scale parameter $\sigma$~\cite{Baccelli}.
\end{itemize}
Note that $X_i(t+\dd t)$, i.e., the belief of agent $i$ at time $t+\dd t$,
depends on her personality $P_i$ and the current
agent's belief. In other words, the temporal evolution of agents' beliefs $
\{ X_i(t),\; i\in \Uc\}$ is Markovian over $\Zc^U$, where $\Zc =\Pc \times \Xc$ is a
bi-dimensional continuous space.

\subsection{From a discrete to a continuous system model}
Given that $\Pc$ and $\Xc$, hence $\Zc$, are continuous spaces, we
define the empirical probability measure,
$\rho^{(U)}(\dd p, \dd x, t)$, over the belief-personality space $\Zc$
at time $t$, as:
\begin{equation}
  \rho^{(U)} (\dd p, \dd x, t ) =\frac{1}{U}\sum_{i \in \Uc}
  \delta_{(P_i,X_i(t ))}(\dd p, \dd x)\,.
  \label{eq:rho_U}
\end{equation}
In the above expression, $\delta_{(P_i,X_i(t ))}(\dd p,\dd x)$ is the
Dirac measure centered at $(P_i,X_i(t ))$. i.e.,
  $\delta_{(P_i,X_i(t ))}(\dd p,\dd x)$ represents the mass
  probability associated with opinion $X_i(t )$ of agent $i$, which
  has personality $P_i$. Note that in~\eqref{eq:rho_U} agents are
seen as particles in the continuous space $\Zc$, moving along the
opinion axis $x$.  Our goal is to describe the evolution of
$\rho^{(U)}(\dd p, \dd x, t)$.  To this end, we perform an asymptotic
analysis by considering the number of agents to grow to
  infinity, i.e., $U \to \infty$. In this case, agents become a
  continuous fluid of particles characterized by personality $p$ and
  opinion $x$ with $(p,x)\in \Zc$.
  In particular, similarly to \cite{Garnier}, we apply the mean-field theory,
  according to which, the effect of all other agents on any given
  agent can be represented by a single average effect. So doing, we
  can exploit the results
  in~\cite{gartner1988mckean,dawson1983critical} and state that, as
  $U\to \infty$, $\rho^{(U)} (\dd p, \dd x, t)$ converges in law to
  the asymptotic distribution $\rho(p, x, t)$, provided that
  $\rho^{(U)} (\dd p, \dd x, 0)$ converges in law to $\rho(p, x,
  0)$. Also, $\rho(p, x, t)$ can be obtained from the following
  non-linear Fokker-Planck (FP)
  equation~\cite{gartner1988mckean,dawson1983critical}: 
\begin{equation} \label{eq:fokker-planck}
\frac{\partial \rho(p, x, t)}{\partial t} = -\frac{\partial [\mu_x(p,x,t,\rho) \rho(p,x, t)]}{\partial x}+ 
 \frac{\sigma^2}{2}\frac{\partial^2\rho(p, x, t) }{\partial x^2}\,.
\end{equation}
In \eqref{eq:fokker-planck},  $\mu_x(p,x,t,\rho)$ is defined as the instantaneous average speed
along axis $x$ of a generic agent located at position $(p,x)$ at time
$t$ (i.e., of an agent with personality $p$ and belief $x$). 
Such instantaneous average speed is given by: 
\begin{equation}
\label{eq:mu_x} 
\mu_x(p,x,t,\rho) =\lim_{\Delta t\to 0} \frac{\EE[X(t\mathord{+}\Delta t)\mid
  P=p,X(t)=x]\mathord{-}x}{\Delta t} \,.
\end{equation}
From \eqref{eq:x_i} and considering that
the Brownian motion process $W(t)$ has zero drift, we write:
\begin{eqnarray}
\label{eq:mu_x2}
\mu_x(p,x,t,\rho) 
&=& \bar{\alpha}(p)\int_{\Zc}
\zeta(|x'\mathord{-}x|,p,p') (x'\mathord{-}x) \nonumber \\ 
&&\cdot \rho(p',x',t) \dd p'\dd x' +\alpha(p) [u(p)\mathord{-}x]
\end{eqnarray}
where both $u(p)$ and $\alpha(p)$ are assumed to be continuous
functions in $p$ and $\zeta(|x'-x|,p,p')$ to be continuous with
respect to its second and third arguments.  Note that, in the RHS of
the above expression, agent interactions are represented by the
integral over $\Zc$ instead of the sum over the set of agents $\Uc$.

In the following, we analyze the system dynamics by solving the above
FP equation in terms of $\rho(p, x, t)$ so as to obtain the distribution of agents over $\Zc$. To this end, we wish
to emphasize that the following properties hold with regard to the
system stability:
\begin{itemize}
\item[{\em (i)}]
When $\Xc=\mathbb{R}$ and 
 $\zeta(|x'-x|,p,p')=1$, $\forall
x,x'\in \Xc$ and $\forall p,p' \in \Pc$, it has been shown \cite{Tempo} that
whenever $\alpha(p)>0$ $\forall p$, the system is stable (i.e., 
beliefs do not drift to infinite); 
\item[{\em (ii)}] More in general, in Appendix A  
we show that the Markovian process defined by~\eqref{eq:x_i}
and, hence, by~\eqref{eq:fokker-planck}, is ergodic when 
$\inf_{p\in \Pc} \alpha(p)>0$.
Thus, the empirical distribution of beliefs $\{X_i(t), i\in \Uc\}$, 
$\rho^{(U)} (\dd p, \dd x, t)$, converges in
law to a limiting distribution for $t\to \infty$, $\rho^{(U)} (\dd p,
\dd x)$, 
which is unique and
independent of the initial condition;   
\item[{\em (iii)}] As a consequence of the fact
that $\rho^{(U)} (\dd p,
\dd x)$ converges to $\rho(p, x, t)$ as $U\to \infty$,   also the asymptotic
distribution $\rho(p, x, t)$ admits a unique limit for
$t\to \infty$, $\rho(p, x)$, independently from the initial condition.  
Note that {\em this limit
can be found as the unique stationary solution
of~\eqref{eq:fokker-planck}}. 
\end{itemize}
In light of the above observations, in the following one of our main
objectives is to find a stationary solution of the above FP equation.

\section{Stationary analysis of the FP equation\label{sec:bounded}}
In this section, we analyze the stationary solution of the FP equation
in \eqref{eq:fokker-planck}. Specifically,
\begin{itemize}
  \item we start with the most general scenario
and we show that such solution corresponds to the fixed point of a
properly defined operator (Sec. \ref{subsec:general});
\item then we deal with the case of unbounded confidence in
  Sec. \ref{subsec:unbounded}, where we provide an alternative,
  simpler, expression for the stationary solution. This allows us to
  derive, under mild additional assumptions, a closed-form expression
  for the stationary solution of the FP equation;
\item in Sec. \ref{subsec:bounded}, we focus on the case where bounded
  confidence holds. We propose an iterative procedure
  and prove that it actually converges to the stationary solution of
  the FP equation in some relevant cases.
\end{itemize}

\subsection{Stationary analysis under general conditions}\label{subsec:general}
By definition, in stationary conditions $\rho(p, x, t)$ is constant
over $t$, i.e.,   $\frac{\partial }{\partial t} \rho(p, x, t) = 0$.
In such a case, we drop the time dependence from the symbols $ \rho(p, x, t)$ and
 $\mu_x(p,x,t,\rho)$. 
Then the stationary solution can be found by setting to zero the
LHS of~\eqref{eq:fokker-planck}  
and integrating the resulting equation, i.e., 
\begin{equation} \label{eq:C}
C=   -\mu_x(p,x,\rho) \rho(p, x) + 
 \frac{\sigma^2}{2}\frac{\partial }{\partial x}  \rho(p, x)  
\end{equation}
where $C$ is a constant that can be determined by imposing boundary conditions.
Observe that, in order to be a solution of the FP equation, $\rho(p,x)$ must be
twice differentiable with respect to $x$, $\forall p\in \Pc$, with
continuous derivatives at every point of the domain $\Xc$. Moreover, we assume
$\rho(p,x)$ to be continuous at every point $(p,x)\in \Zc$. 

We first focus on the solution of~\eqref{eq:C} when $\Xc=\mathbb R$.
In such a case, we impose the following conditions: (i)
$\int_{\Pc}\int_{\Xc} x^2 \rho(p,x) \dd x\dd p <\infty$, i.e., the
second moment of the steady-state belief distribution to be finite
(note that this obviously implies that $\rho(p,x) \to 0$ faster than
$1/|x|^3$ as $x \to \infty$), and (ii)
$ \int_{\Xc} x^2 \frac{\partial{\rho(p,x)}}{{\partial x}} \dd x<
\infty$ $\forall p\in \Pc$.  For short, we denote by
$\mathbb{C}^{2}_x(\Pc, \Xc, x^2)$ the space of distributions that
satisfy the above conditions, and by $\mathbb{L}_{1}(\Pc, \Xc, x^2)$
the class of summable functions satisfying (i) only.

Since~\eqref{eq:C} must hold for any $x$ and, hence, also as $x\to
\infty$, we must have $C=0$.  Indeed, by \eqref{eq:mu_x2},
$\mu_x(p,x,\rho)<C_1|x|$ for a properly defined $C_1$, thus, for  $\rho(p,x) \in \mathbb{C}^{2}_x(\Pc, \Xc, x^2)$, both terms on the RHS of \eqref{eq:C} tend to 0 as $x\to
\infty$. By setting $C=0$ in~\eqref{eq:C}, 
we have 
\begin{equation} \label{eq:fokker-planck-stat-3}
\mu_x(p,x,\rho) \rho(p,x)  =     \frac{\sigma^2}{2}\frac{\partial }{\partial x}  \rho(p,x)\,. 
\end{equation}
For every $p\in \Pc$,~\eqref{eq:fokker-planck-stat-3} can be formally
solved by dividing by $\rho(p,x)$ (under the assumption that
$\rho(p,x)>0$ at every point) and integrating both sides, as:
\begin{equation} \label{eq:general-solution}
\rho(p,x)= K(p) \exp\left(\frac{2}{\sigma^2}\int_{0}^x \mu_x(p,y,\rho) \dd y\right)\rho_0(p) 
\end{equation}
where $K(p)$ is a normalizing function such that 
\[ K(p)\int_\Xc \exp\left(\frac{2}{\sigma^2}\int_{0}^x \mu_x(p,y,\rho) \dd y\right) \dd x=1 \,,\]
and $\rho_0(p)$ is the distribution  representing the agent personality density, i.e.,
\begin{equation}\label{eq:rho0} 
\rho_0(p)=\int_\Xc \rho(p,x) \dd x= \int_\Xc \rho(p,x,0) \dd x\,.
\end{equation} 
Note that the~\eqref{eq:rho0} descends from the fact that agents'
personality exhibits no dynamics; 
in the following we assume $\rho_0(p)$  to be a continuous and bounded function. 

Importantly, the solution in \eqref{eq:general-solution} is an
{\em implicit} expression for $\rho(p,  x)$, since $\mu_x$ actually
depends on $\rho$.  
Also, we remark that when $\Xc$ is compact, \eqref{eq:general-solution} is the solution
of~\eqref{eq:C} under reflection boundary conditions which are
obtained by imposing that no mass crosses the boundaries $\partial\Xc$, i.e.,
$\mu(p,x)\rho(p,x)-\frac{\sigma^2}{2} \frac{\partial\rho(p,x)}{\partial
  x}\mid_{x\in \partial \Xc}=0$. 

By replacing the expression of $\mu_x(p,x,\rho)$ in
\eqref{eq:general-solution} with~\eqref{eq:mu_x2}, we get:
\begin{eqnarray} \label{eq:rho-2}
\rho(p,x)
&=& K(p) \exp\left(
  \frac{2 \bar{\alpha}(p)}{\sigma^2}\int_{0}^x \int_{\Zc}
  \zeta(|y'\minus y|\mathord{,}p\mathord{,}p') \right . \nonumber \\
&&\qquad  \bigg . \cdot (y'\minus y)  \rho(p'\mathord{,}y') \dd
  p'\dd y' \dd y\bigg)  \nonumber \\ 
&&\qquad  \cdot \exp\left(-\frac{\alpha(p) [x-u(p)]^2}{\sigma^2}\right) \rho_0(p) \,.
\end{eqnarray}
The solution in~\eqref{eq:rho-2} corresponds to the fixed point of the operator  
$\Ac\{\rho\}: \mathbb{L}_{1}(\Pc, {\mathbb R}, x^2)  \to \mathbb{C}^{2}_x(\Pc, {\mathbb R}, x^2)$,
  defined as: 
\begin{equation} \label{eq:A}
\Ac\{\rho\}= K(p) \exp\left(\frac{2}{\sigma^2}\int_{0}^x \mu_x(p,y,\rho) \dd y\right)\rho_0(p) \,.
\end{equation}
Note that the image of $\Ac\{\rho\}$ belongs to
$\mathbb{C}^{2}_x(\Pc, {\mathbb R}, x^2)$. Indeed, by construction,
$\mu_x(p,x,\rho)$ (see \eqref{eq:mu_x2}) is continuous and
differentiable with continuous derivative with respect to $x$ over
$\Xc$, for any choice of a summable $\rho(p,x)$ (we recall that
$\zeta(|x-x'|,p,p')$ is assumed smooth). This implies that the image
of any summable function $\rho(p,x)$ through $\Ac\{\rho\}$ is
necessarily continuous and twice differentiable with respect to $x$.
Similarly, for $x\to \pm \infty$, it can be shown that $\Ac\{\rho\}$,
along with its first derivative, exhibits a tail that goes to zero
faster than exponentially, since $\rho(p,x) \to 0$ faster than $1/x^3$
(implied by the fact that
$\rho(p,x)\in \mathbb{L}_{1}(\Pc, {\mathbb R}, x^2) $).

Also, the fixed point of operator $\Ac\{\rho\}$, being a point of the image of
$\Ac\{\rho\}$, must belong to $\mathbb{C}^{2}_x(\Pc, {\mathbb R},
x^2)$.  As an immediate consequence, by construction, any solution of the
original stationary FP equation is necessarily a fixed point of~\eqref{eq:rho-2}
and vice versa.  At last, observe that the argumentation at the end of
Section \ref{sec:model} about the existence and uniqueness of the stationary
distribution for the considered dynamical system (and the associated FP
equation), guarantees that a unique fixed point (satisfying $\rho(p,x)\ge 0$) of
the previous operator always exists.

\subsection{The case of unbounded confidence\label{subsec:unbounded}}
We now consider the simpler case in which $\Xc=\mathbb{R}$ and 
$\zeta(|x'-x|, p,p')=\zeta(p,p')$. The latter reflects  
the scenario in which the interactions between
agents are effective regardless of the difference between their beliefs (no
bounded confidence). In this case, we show that an expression for the steady-state
distribution $\rho(p,x)$ can be obtained in terms of the solution of a linear
second-type Fredholm integral equation.  In particular, when we consider
that the influence of interactions on the user's belief is modeled as $\zeta(
p,p')=\zeta_1(p)\zeta_2(p')$, i.e., according to the well-known
gravity model\footnote{In this case, users attract each other
  according to their personality value similarly to what happens to
  two different masses in the gravity model.}, a closed-form expression for  $\rho(p,x)$ is
obtained.

We start by rewriting the integrals in~\eqref{eq:rho-2}, as:
\begin{eqnarray}
I&=&\int_0^x\int_{\Zc} \zeta(p,p')(y'\minus y)   \rho(p',y') \dd y' \dd p'\dd y \nonumber\\ 
&=&\int_0^x \int_{\Pc}\int_{\Xc}y'  \zeta(p,p' )\rho(p',y') \dd y' \dd p'\dd y\nonumber \\
&&\qquad -\int_0^x y \int_{\Pc}\zeta(p,p')\rho_0(p') \dd p'\dd y  \nonumber \\ 
&=&\int_0^x \left[\beta(p) -y \eta(p)\right]\dd y    \nonumber \\
&=&-\frac{\eta(p)}{2}[x-\phi(p)]^2 +\frac{\eta(p)}{2}\phi(p)^2\label{eq:I}
\end{eqnarray}
where we used the definition of $\rho_0(p)$ given in~\eqref{eq:rho0} and we defined:
\begin{equation} 
\eta(p)= \int_{\Pc}\zeta(p,p')\rho_0(p')\dd p'\,, \label{eq:etap}
\end{equation}
\begin{equation}
 \beta(p)=  \int_{\Pc}   \zeta(p,p' ) \int_{\Xc}  y'  \rho(p',y') \dd y'  \dd p'\,,
\label{eq:betap}
\end{equation}
and $\phi(p)= \beta(p)/\eta(p)$  (we assume $\inf_{p\in \Pc} \eta(p)>0$).
By substituting~\eqref{eq:I} in~\eqref{eq:rho-2} and after some
algebra, we get the distribution $\rho_\phi(p,x)$, which, given $p$, turns out to
be Gaussian with mean $m(p)$. Indeed, we obtain:
\begin{equation}\label{rhoexpression}
\rho_\phi(p,x) = \sqrt{\frac{w(p)}{\pi\sigma^2}} \exp\left(-\frac{w(p)\left[x\minus
    m(p) \right]^2}{\sigma^2}\right) \rho_0(p)
\end{equation}
where $w(p) = \alpha(p)+\bar{\alpha}(p)\eta(p)$, and
\begin{equation}\label{thetap}
m(p)= \frac{\alpha(p) u(p)+ \bar{\alpha}(p) \eta(p) \phi(p)}{w(p)} \,.
\end{equation}
Intuitively, $m(p)$ can be seen as the convex combination of the
prejudice and $\phi(p)$, with the latter representing
the (normalized) impact of social interaction on the opinion of agents
with personality $p$.
In~\eqref{rhoexpression} we added the subscript $\phi$ to $\rho(p,x)$ in order to
stress the dependence on $\phi$. Therefore, the image of the
operator $\Ac\{\rho_\phi\}$ lies within the space of functions defined
by~\eqref{rhoexpression}.  It is then licit to expect that also the fixed point
of operator $\Ac\{\rho_\phi\}$, i.e., $\rho_\phi^*(p,x)$, is in the
form~\eqref{rhoexpression} for a specific function $\phi^*(p)$. Now, in order to find
$\phi^*(p)$, we must impose that $\rho_\phi^*(p,x)=\Ac\{\rho_\phi^*(p,x)\}$.  After some
algebra, it turns out that $\phi^*(p)$ satisfies the following inhomogeneous
Fredholm integral equation of the second kind~\cite{kolmogorov2012functions}:
\begin{equation}\label{eq:beta-star2}
\phi^*(p)= h(p) + \int_{\Pc}\Gamma(p,p')\phi^*(p') \dd p'
\end{equation}
with $h(p)\mathord{=}\int_{\Pc} \frac{\zeta(p,p')\rho_0(p') \alpha(p') u(p')}
 {\eta(p) w(p')} \dd p' $ and $\Gamma(p,p')=\frac{ \zeta(p,p' )
 \rho_0(p')\bar{\alpha}(p')\eta(p')}
 {\eta(p)w(p')}$.

 Under the assumption that $\Gamma(p,p')$ is bounded over its domain, the
 existence and uniqueness of the solution of~\eqref{eq:beta-star2} is guaranteed whenever the associated homogeneous Fredholm operator
 $\Bc\{V(p)\} = W(p)=\int\Gamma(p,p') V(p')\dd p'$ does not admit
 non-zero fixed points (i.e., function $V(p)\neq 0$ satisfying
 $V(p)=\Bc\{V(p)\}$) \cite{kress1989linear}.  In
 our case, by direct inspection, it can be verified that 
$\sup_p \int_\Pc \Gamma(p,p') \dd p' <1$, and therefore $\forall p\in
\Pc$:
\begin{eqnarray} 
|V(p)| \! &=& \!\Big|\int_\Pc \Gamma(p,p') W(p') \dd p'\Big| \le \int_\Pc
\Gamma(p,p')| W(p')| \dd p' \nonumber\\
&\le & \int_\Pc \Gamma(p,p') \sup_{p'} | W(p')| \dd q 
< \sup_{p'} | W(p')| \,. 
\end{eqnarray}  
We conclude that  $\sup_p |V(p)|
 <\sup_p |W(p)|$ whenever $\sup_p |W(p)|>0$ and, therefore, the equation
 $\Bc\{V(p)\}= V(p)$ does not admit non-zero solutions.

Also, under the previous assumption, the unique solution of the Fredholm
equation in \eqref{eq:beta-star2}
can be expressed in terms of the so-called Kernel resolvent $H(p,p')$ as:
\begin{equation}\label{eq:sol-Fredholm}
 \phi^*(p)=h(p)-\int_{\Pc}H(p,p')h(p') \dd p' \,.
\end{equation}
In the simplest case (i.e., when the $\mathbb{L}_2$-norm of the kernel is
smaller than 1: $\int_{\Pc}\int_{\Pc}\Gamma^2(p,p')\dd p \dd p'<1$), $H(p,p')$
can be expressed as $H(p,p')=\sum_{m=0}^\infty \Gamma_m(p,p')$, being
$\Gamma_m(p,p')$ the $m$-th iterated kernel satisfying the following
recursion
\cite{kress1989linear,kolmogorov2012functions}: $
\Gamma_m(p,p')=\int_{\Pc}\Gamma_{m-1}(p,w)\Gamma(w,p')\dd w$.  In such a case,
$H_n(p,p')=\sum_{m=0}^n \Gamma_m(p,p')$ converges uniformly to $H(p,p')$ over
the whole domain { \cite{kolmogorov2012functions,kress1989linear}}.  Therefore,
the sequence $\rho_{\phi,n}^*(p)$, where
$\phi_n^*(p)=h(p)-\int_{\Pc}H_n(p,p')h(p') \dd p'$, converges uniformly to
$\rho_\phi^*(p)$ as immediate consequence of the  theorem below.
\begin{teorema} \label{teo_lin_cont_extend}
Whenever $\phi(p)\to \phi^*(p)$ uniformly over $\Pc$, then
$\rho_{\phi}(p,x)\to \rho_{\phi}^*(p,x)$ uniformly over $\Zc$ 
 under the assumptions: $\sup_{p}\rho_0(p)<\infty$ and $\inf_p \alpha(p)>0$. 
\end{teorema}
{\em Proof:} See Appendix B.

It is worth noticing that when the
$\mathbb{L}_2$-norm of the kernel is greater than 1, the resolution kernel
$H(p,p')$ can still be expressed as a ratio of two
series~\cite{kolmogorov2012functions,kress1989linear}.  Also, several
efficient numerical techniques have been developed for the solution of Fredholm
equations.

Finally, we focus on the case where
$\zeta(p,p')= \zeta_1(p)\zeta_2(p')$.  We stress that $\zeta_1(p)$ essentially measures
to what extent the belief of agents with personality $p$ are
influenced by other agents' beliefs; $\zeta_2(p')$ instead measures
the degree of influence exerted by agents with personality $p'$ on all
other agents. In such a case the above steady-state analysis greatly
simplifies.

Indeed, $\eta(p)$ and $\beta(p)$ (in~\eqref{eq:etap}
and~\eqref{eq:betap}, respectively) can be 
expressed in terms of the unknown scalar parameter $\eta$ and $\beta$, respectively:
\[
\eta(p)= \zeta_1(p)\int_{\Pc}\zeta_2(p')\rho_0(p')\dd p'= \zeta_1(p) \eta
\]
with $ \eta= \int_{\Pc}\zeta_2(p')\rho_0(p') \dd p'$, and
\[ \beta(p)= \zeta_1(p) \int_{\Zc}  y \zeta_2(p' )  \rho(p',y) \dd y  \dd p' = \zeta_1(p) \beta \]
 with $\beta =\int_{\Zc} y \zeta_2(p') \rho(p',y) \dd y \dd p'$.
 Hence, $\phi(p)= \frac{\beta(p)}{\eta(p)}=\frac{\beta}{\eta} =\phi^*$ becomes a
 constant, .e., it does not depend on the agent's
   personality $p$.
By casting~\eqref{eq:beta-star2} in this specific case, we can
 express $\phi^*$ as:
\begin{equation}\label{eq:phi-star}
 \phi^* =  \frac{1}{\eta}\frac{\int_{\Pc}\frac{\zeta_2(p') \rho_0(p')\alpha(p') u(p')} {\alpha(p') + \bar{\alpha}(p')\zeta_1(p')\eta }\dd p' }
 { 1-  \int_{\Pc} \frac{\zeta_2(p') \rho_0(p')\bar{\alpha}(p')\zeta_1(p') }
   {\alpha(p') + \bar{\alpha}(p')\zeta_1(p')\eta  }  \dd p'} \,.
\end{equation}
In conclusion, by replacing the above expression into \eqref{thetap}
and using~\eqref{rhoexpression}, 
we obtain a closed-form expression for the steady-state distribution of
the agents' beliefs. 

We remark that this result can be generalized to the case of any
$\zeta(p,p')$ that can be expressed as: $\zeta(p,p')=
\sum_{i=1}^K\sum_{j=1}^Ka_{ij} \zeta^{(i)}(p)\zeta^{(j)}(p')$, for
some $K\in \mathbb{N}$ and with $\zeta^{(k)}(p)$ being orthonormal
functions. The class including such functions is dense in
$\mathbb{L}_2$-norm, hence every $\zeta(p,p')\in \mathbb{L}_2$ can be
approximated with an arbitrary degree of accuracy (with respect to the
$\mathbb{L}_2$-norm) by a function belonging to such class.

\subsection{The case  of bounded confidence}\label{subsec:bounded}
The study of belief dynamics in the presence of bounded confidence
is challenging, due to the non-linearity emerging in the 
interactions between agents. 
Thus, in this case we propose an iterative procedure to obtain the unique
stationary distribution  of the FP equation.

\subsubsection{A successive approximation methodology}\label{sec:succapprox}
The solution of the FP equation, i.e., the fixed point of the 
operator $\Ac\{\rho\}$ in~\eqref{eq:A}, can be, in principle, obtained by  a successive
approximation method, provided that the sequence of functions we use converges.
Let us then define $k$ as the iteration index and the following sequence of
functions:
\begin{eqnarray}
\mu_x^{(k)}(p,x,\rho^{(k)})&\hspace{-3mm}\triangleq\hspace{-3mm}& \bar{\alpha}(p)\int_{\Zc}\zeta(|x'\minus x|,p,p') (x'\minus x)\nonumber \\
&&\qquad \cdot \rho^{(k)}(p',x') \dd p'  \dd x' \label{eq:muxk} \\
\rho^{(k+1)}(p,x) &\triangleq & K^{(k)}(p) \exp\left(-\frac{\alpha(p)}{\sigma^2} [x-u(p)]^2\right) \nonumber \\
& &\cdot \exp\left(\int_{0}^x 
 \frac{2\mu_x^{(k)}(p,y,\rho^{(k)})}{\sigma^2}\dd y\right) \rho_0(p) \nonumber \\
\end{eqnarray}
where $K^{(k)}(p)$ is the normalization constant at the $k$-th iteration.
The iterative procedure starts with $\mu_x^{(0)} = 0$.

Note that, if {\em (i)} $\rho^{(k)}(p,x)$ converges to a point $\rho^*(p,x)$
under some convergence criterion and {\em (ii)} the operator $\Ac\{\rho\}$ is
continuous at $\rho^*(p,x)$ under the same convergence criterion, then
$\rho^{(k)}(p,x)$ converges to the fixed point of operator $\Ac\{\rho\}$ (which
we know to exist and to be unique)~\cite{kolmogorov2012introductory}.  In
particular, convergence to the unique fixed point is exponentially fast whenever
the operator $\Ac\{\rho\}$ is a contraction map \cite{kolmogorov2012introductory}.

Proving properties for the operator $\Ac\{\rho\}$ in the general case
of $\Xc=\RR$ is challenging. However, in the case of practical
relevance where $\Xc$ is compact and under reflection boundary
conditions and weak interactions between agents, we show below
(Sec. \ref{subsec:compact-weak}) that $\Ac\{\rho\}$ is a contraction
map, thus our successive approximation procedure converges to the
solution of the FP equation.  Furthermore, we prove the
continuity of the operator $\Ac\{\rho\}$ (i.e., condition {\em (ii)}
above) under fairly general conditions, i.e., even when we relax the
assumption about the strength of interactions, or $\Xc=\RR$ (see
Appendices C and D).
Thanks to this result, in these latter cases
we can consistently apply our iterative procedure whenever we have (numerical)
evidence that $\rho^{(k)}(p,x)$ converges (i.e., also condition {\em (i)} above
is met).

\subsubsection{Compact $\Xc$ and limited agents'
  interactions\label{subsec:compact-weak}} 
Here we consider the 
belief space, $\Xc$, to be compact and the strength of interactions, i.e.,
$\zeta(|x'-x|, p,p')$, to be relatively small. In this case, 
the following theorem holds.
\begin{teorema} \label{th:contraction}
Whenever $\Xc$ is a compact set, the operator $\Ac\{\rho\}$ is a contraction map
operating over probability distribution functions with respect to the
$\mathbb{L}_1$-norm if:
\begin{equation}\label{eq:cond-th:contraction} 
S_\zeta S_x  X_0  <\frac{\sigma^2}{8}
\end{equation}
where 
$S_\zeta = \sup_{|x'-x|,p,p'}
\zeta(|x'-x|,p,p')$, $\zeta(|x'-x|,p,p')=0$ for any $|x'-x|>X_0$, and
$S_x=\sup_{x\in \Xc}{|x|}<\infty$.
\end{teorema}

{\em Proof:} See Appendix D.

By virtue of the above result, under \eqref{eq:cond-th:contraction} it is
guaranteed that our iterative procedure converges to the only fixed point of the
operator $\Ac\{\rho\}$ in the $\mathbb{L}_1(\Pc,\Xc)$-space. Additionally, as
observed before, the fixed point of the operator $\Ac\{\rho\}$ is twice
differentiable, which therefore represents a suitable solution of the FP
equation.~\footnote{Condition \eqref{eq:cond-th:contraction}, by no means,
  should be considered necessary for the operator $\Ac\{\rho\}$ to be a
  contraction map.  Indeed, the proof of Th.  \ref{th:contraction} is based on a
  chain of inequalities, some of which may be rather loose in many cases.}
Indeed, recall that the solution of the FP equation must be continuous and twice
differentiable in $\Xc$, and continuous in $\Pc$. 
Alternatively, when the condition stated in \eqref{eq:cond-th:contraction} does
not hold, we can prove the following results:

{\em (1)} the operator $\Ac\{\rho\}$ is continuous at the fixed point
(w.r.t. convergence in $\mathbb{L}_1$-norm), as shown in Appendix C.
Thus, provided that $\rho^{(k)}(p,x)$ converges to $\rho^*(p,x)$, necessarily
$\rho^*(p,x)$ is the fixed point of the operator $\Ac\{\rho\}$;

{\em (2)} under milder conditions with respect to Theorem~\ref{th:contraction},
$\Ac\{\rho\}$ is shown in Theorem \ref{th:contraction-loc} below, to be a
contraction map in an arbitrarily small neighborhood $\Uc$ of the fixed point
$\rho^*(p,x)$. This guarantees that the successive approximation procedure
converges to the operator's fixed point whenever it reaches $\Uc$.
\begin{teorema}\label{th:contraction-loc}
Whenever $\Xc$ is a compact set, the operator $\Ac\{\rho\}$ is a contraction
map, operating over probability distribution functions in a neighborhood of
$\rho^*(p,x)$, if:
\[ S_\zeta   S_x X_0 <\frac{\sigma^2}{2}-\epsilon \qquad \forall \epsilon>0 \,. \]
\end{teorema}

{\em Proof:} See Appendix D.

\section{Results on stationary belief distribution}\label{subsec:results-steady}
Here we present some numerical results obtained by casting our techniques and
the expressions obtained above, considering $\Xc=\mathbb{R}$.
Specifically,
\begin{itemize}
\item in Sec.~\ref{subsec:4.1} we start with the simplest case in
  which beliefs evolve according to an unbounded confidence model and
  $\zeta(p,p')$ can be expressed in product form, i.e.,
  $\zeta(p,p')= \zeta_1(p)\zeta_2(p')$. Under these conditions,
  \eqref{eq:phi-star} along with \eqref{rhoexpression} and
  \eqref{thetap} provide a direct expression for the steady-state
  distribution;
  \item then in Sec.~\ref{subsec:4.2} we move to the more general case of unbounded
  confidence and a generic $\zeta(p,p')$.  The steady-state
  distribution can now be obtained by solving the Fredholm integral
  equation~\eqref{eq:sol-Fredholm} in $\phi^*(p)$;
  \item at last, in Sec.~\ref{subsec:4.3} we
  consider the most challenging scenario where bounded confidence is
  in place. In this case, we evaluate numerically the steady-state
  distribution through the method of successive approximations
  presented in Sec.~\ref{sec:succapprox}.
\end{itemize}

\subsection{Unbounded confidence  with 
$\zeta(|x\minus x'|\mathord{,}p\mathord{,}p')= \zeta_1(p)\zeta_2(p')$\label{subsec:4.1}}
We consider agents' personalities to be distributed over a finite interval
centered around the origin, i.e., $\Pc = [-P,P]$.  Without loss of generality, we
set $P=1$.  In this case we observe that when $\alpha(p)$, $\rho_0(p)$,
$\zeta_1(p)$ and $\zeta_2(p')$ are even functions, while $u(p)$ is an odd
function, we have $\phi^*=0$. Indeed, the numerator of~\eqref{eq:phi-star} is
the integral of an odd function over a domain that is symmetric w.r.t. the
origin.  Thus, the steady-state distribution is given by:
\begin{equation}\label{solution-res}
\rho(p,x) = \sqrt{\frac{w(p)}{\pi\sigma^2}}\exp\left(-\frac{w(p)[x- m(p)]^2}{\sigma^2}\right) \rho_0(p) 
\end{equation}
with $ m(p) = \frac{\alpha(p) u(p)}{w(p)}$. Note that we can easily
obtain the marginal distribution of the beliefs at steady state, $\rho(x)$,
 by integrating $\rho(p,x)$ over $\Pc$. 

In the following, unless otherwise specified, we set $u(p)=p$, i.e.,
we identify the user's prejudice with her personality; hence, the
user's stubbornness, $\alpha(p)$, and the users' mutual influence,
$\zeta(p,p')$, depend on the user's prejudice.  Below we consider two
specific scenarios highlighting the impact of the system parameters
on the belief stationary distribution.

\begin{figure}[htb]
\centering
\includegraphics[width=0.7\columnwidth]{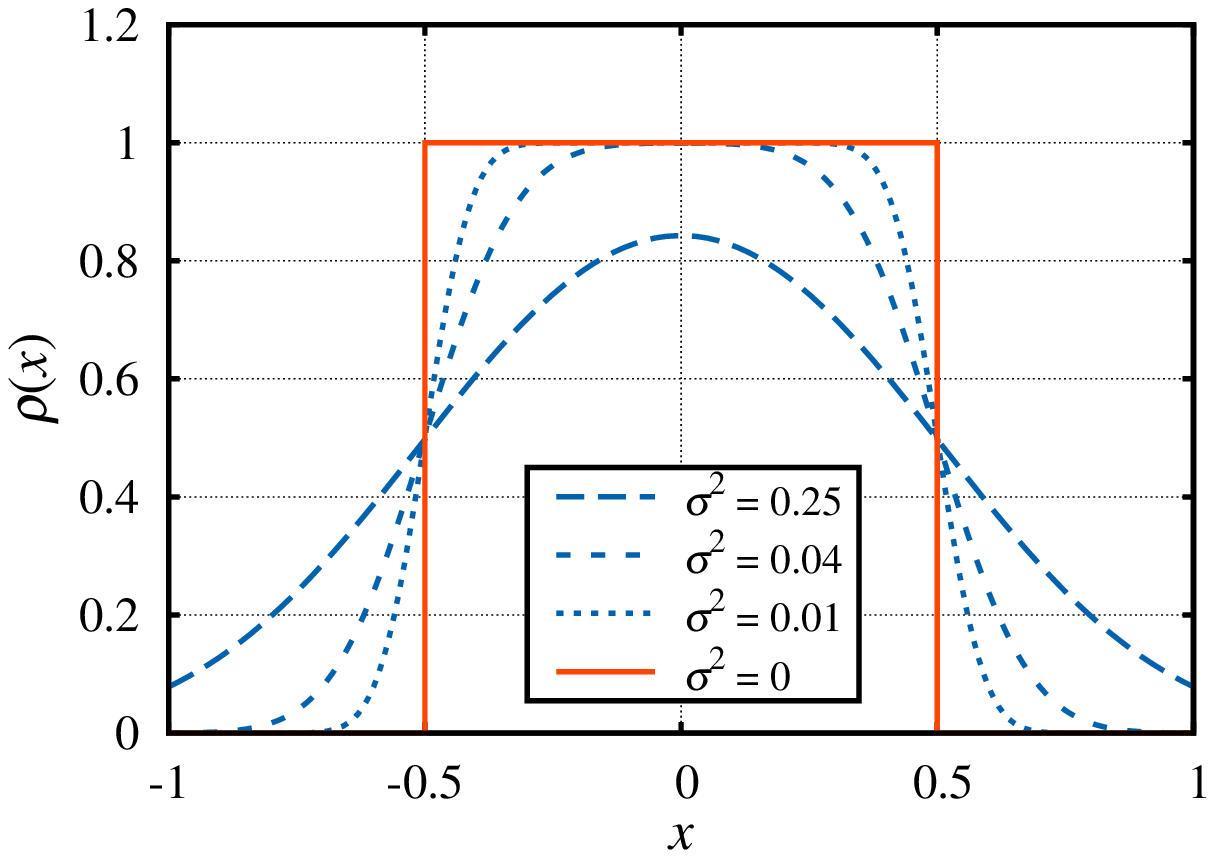}
\includegraphics[width=0.7\columnwidth]{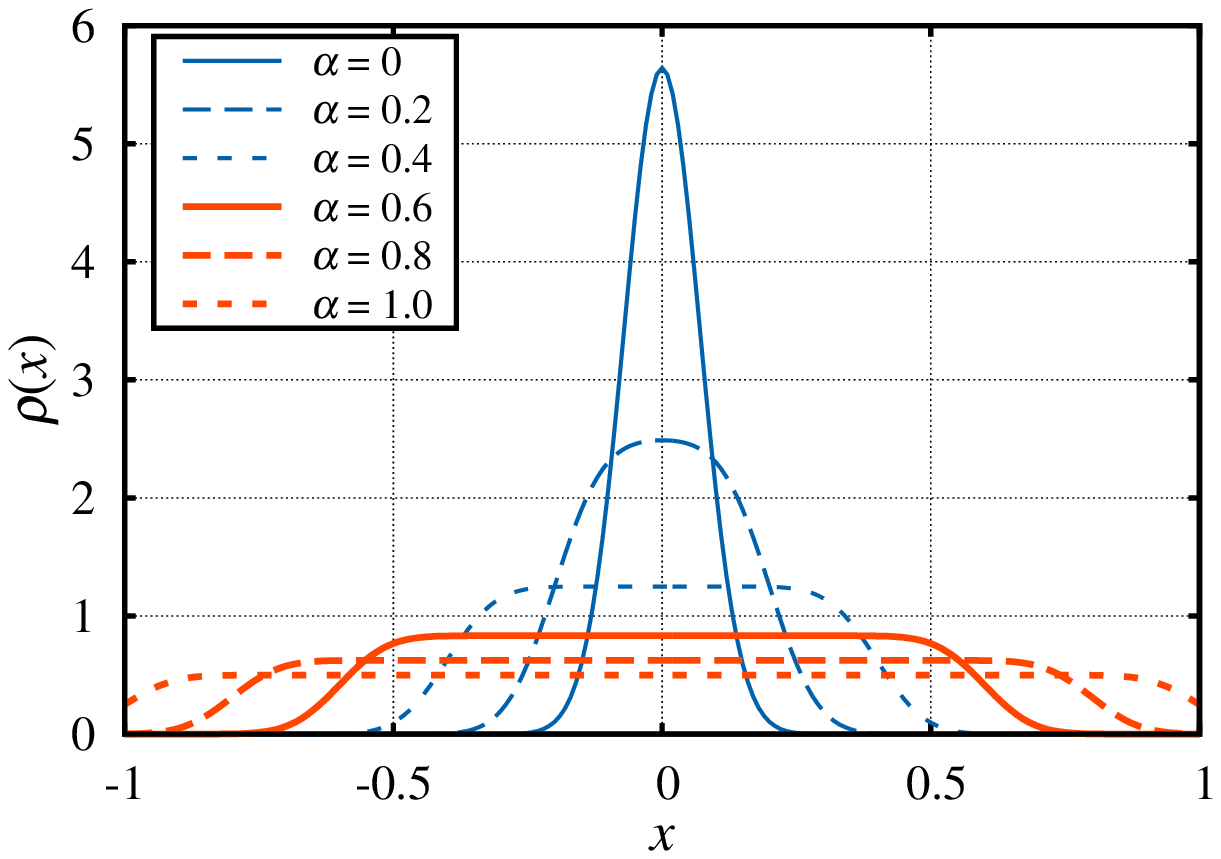}
\caption{Steady-state distribution of beliefs under unbounded
  confidence and in the case of homogeneous agents and constant mutual influence. (Top) Impact of
  the endogenous noise when $\alpha =0.5$. (Bottom) Impact of users'
  stubbornness ($\alpha$) when endogenous noise is constant
  ($\sigma^2 =0.01$).}
\label{fig:effetti_sigma_alpha}
\end{figure}

\begin{figure}[htb]
  \centering
  \includegraphics[width=0.7\columnwidth]{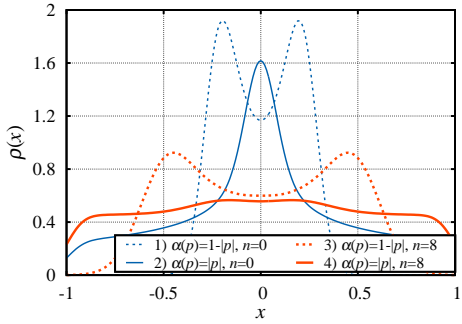}
\caption{Steady-state distribution of beliefs under unbounded confidence
and in the case of inhomogeneous agents and asymmetric mutual influence, different levels of stubbornness, and $\sigma^2 =0.01$.}
\label{fig:effetto_corna}
\end{figure}

\smallskip \noindent {\bf Homogeneous agents and constant mutual
  influence.}  In this case we consider that all agents exhibit the
same level of stubbornness ($\alpha(p)=\alpha$) and the strength of
their interaction does not depend on their beliefs or personality
($\zeta_1(p)=\zeta_2(p')=1$).  Also, we assume a uniform distribution
of agents' personality over $\Pc=[-1,1]$, i.e., $\rho_0(p)=1/2$,
$p\in \Pc$.

In this case, from~\eqref{eq:etap}, we get $\eta(p)=1$ and $w(p)=1$,
and, from~\eqref{solution-res} we obtain the following simple
closed-form expression for the steady-state distribution:
\[ \rho(x) = \frac{1}{4\alpha} \left[ {\rm
      erf}\left(\frac{\alpha+x}{\sigma}\right)+ {\rm
      erf}\left(\frac{\alpha-x}{\sigma}\right) \right]. \] The results
are presented in the plots in Fig.~\ref{fig:effetti_sigma_alpha},
which show the significant role played  on $\rho(x)$ 
by, respectively, the noise variance $\sigma^2$ and the parameter
$\alpha$. 

We observe that when the impact of the endogenous noise is negligible
($\sigma^2 \to 0$), the belief steady-state distribution becomes
uniform as the personality distribution, although more concentrated
around 0.  As $\sigma^2$ increases, the noise process tends to
dominate over the agents' interactions, and the belief distribution
tends to become smoother (see
Fig.~\ref{fig:effetti_sigma_alpha}(top)). We remark that, since our
expressions cannot be directly applied when $\sigma^2=0$, this case
has been extrapolated as limiting trajectory for the steady-state
distribution when $\sigma^2 \to 0$.

Then in Fig.~\ref{fig:effetti_sigma_alpha}(bottom) we set
$\sigma^2=0.01$ and let the level of stubbornness, $\alpha$, vary.
We note that in the case of highly fickle agents (i.e., $\alpha \to 0$),
the belief distribution tends to concentrate around 0, and consensus is not
reached only because of the noise (note that also the curve for $\alpha = 0$ is obtained as
a limit). Indeed, by letting both $\alpha$ and $\sigma^2$ tend to 0, it
can be easily shown that the steady-state distribution tends to a Dirac measure
centered in 0.  As $\alpha$ increases, agents become less and less sensitive
to others' beliefs, weighting more their own prejudice. As a result, beliefs are
increasingly spread out as $\alpha$ grows. For $\alpha=1$, all agents
are stubborn (i.e., completely insensitive to others' beliefs), thus the
steady-state distribution closely resembles the one of the prejudice, and the
differences are only due to the presence of noise.

\smallskip \noindent {\bf Inhomogeneous agents and asymmetric mutual
  influence.}  We now consider a more complex scenario where both
agents' stubbornness and their mutual influence depend on the agents'
personality.  In particular, we recall that $u(p)=p$ and we set
$\zeta_1(p)=1$ and $\zeta_2(p') =(\alpha(p'))^n$, i.e., the more
stubborn an agent is, the higher her influence on others' beliefs.

Fig.~\ref{fig:effetto_corna} depicts $\rho(x)$ in the following scenarios:
\begin{enumerate}
\item $\alpha(p) = 1-|u(p)|=1-|p|$  and $n=0$;
\item $\alpha(p) = |u(p)|=|p|$ and  $n=0$;
\item $\alpha(p) = 1- |u(p)|=1-|p|$ and   $n=8$;
\item $\alpha(p) = |u(p)|=|p|$ and  $n=8$.
\end{enumerate}
In scenarios 1) and 3), stubborn agents have a ``neutral
value'' as prejudice (i.e., prejudice equal to 0), while in scenarios
2) and 4) stubborn agents have extremal prejudices (i.e., equal to 1
or $-1$).
Moreover, in scenarios 1) and 2) all agents exert the same (high)
degree of influence (i.e., $\zeta_2(p')=1$).
Instead, in scenarios 3) and 4), agents exert different degrees
of influence, with  stubborn agents being the top influential ones and most of 
the remaining agents exerting marginal influence on other agents.

As expected, in the case in which stubborn users have neutral belief
(dotted lines in Fig.~\ref{fig:effetto_corna}), they tend to attract
other agents, shifting their beliefs toward the center. Such an action
is not completely successful since the other agents are still
conditioned by their prejudice and the probability mass corresponding
to stubborn agents is limited.  Comparing scenario 1) to scenario 3)
(dotted lines, blue and red, respectively), we can observe that in the
former case the mutual influence between fickle agents reinforces the
attractive effect of stubborn agents.  When instead stubborn agents
have extremal beliefs, i.e., in scenarios 2) and 4) denoted by solid
lines in Fig.~\ref{fig:effetto_corna}, the attractive influence on
others exerted by stubborn agents with positive belief, is almost
nullified by the attractive influence exerted by stubborn agents with
negative belief.  Therefore, the vast majority of fickle users tend to
converge toward a neutral belief, due to the mutual attraction between
themselves.  This effect of convergence toward the center tends to
vanish as the mutual attraction between fickle agents becomes weaker
(i.e., $n$ increases).

\begin{figure}[htb]
  \centering
  \includegraphics[width=0.7\columnwidth]{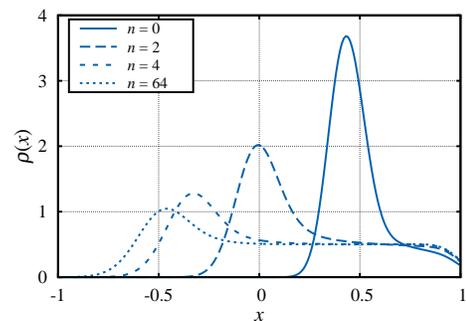}
  \caption{Belief steady-state distribution under unbounded confidence
    and in the case of proximity-based influence. Stubborn users have
    biased positive prejudice, and $\sigma^2=0.01$.}
  \label{fig:effetto_tsunami}
\end{figure}

\subsection{Unbounded confidence with $\zeta(|x,x'|, p,p')= \zeta(p,p')$\label{subsec:4.2}}
Now we move to the more general case in which $\zeta$ cannot be
expressed in product form. In this case, our goal is to assess the
possible effects of the underlying social structure (social graph) on
the belief dynamics. We still consider a scenario in which agents'
personalities are uniformly distributed over $\Pc=[-1,1]$ and
$u(p)=p$.

\smallskip \noindent {\bf Proximity-based mutual influence.}  We set
$\alpha(p)=(p+1)^2/4$, so that stubborn users are those with $p=1$
while fickle users are those with $p=-1$ (hence, they have biased
positive and negative prejudice, respectively).  We compare scenarios
obtained by selecting $\zeta(p,p')= \frac{2}{1+(5|p-p'|)^n}$ for
different values of $n$. This expression for $\zeta(p,p')$ allows us
to study several interesting cases. For $n=0$, it reduces to the case
in which $\zeta$ is constant, namely, $\zeta=1$. As $n$ increases, the
attraction between agents with similar prejudice (i.e., whenever
$|p-p'|<1/5$) tends to become stronger ( $\zeta(p,p')$ for
$|p-p'|<1/5$ saturates to 2 as $n \to \infty$), while the attraction
between agents with different prejudice (i.e., $|p-p'|>1/5$) tends
to vanish. It follows that, although we consider the unbounded case,
users only interact with other agents in their ``proximity''.

Fig.~\ref{fig:effetto_tsunami} shows the results for a decreasing
interaction strength ($n=0, 2,4, 64$). As expected, the underlying
social structure plays a relevant role in the belief dynamics, as it
mitigates the ability of stubborn agents to attract other agents.

\begin{figure}[htb]
\centering
\includegraphics[width=0.7\columnwidth]{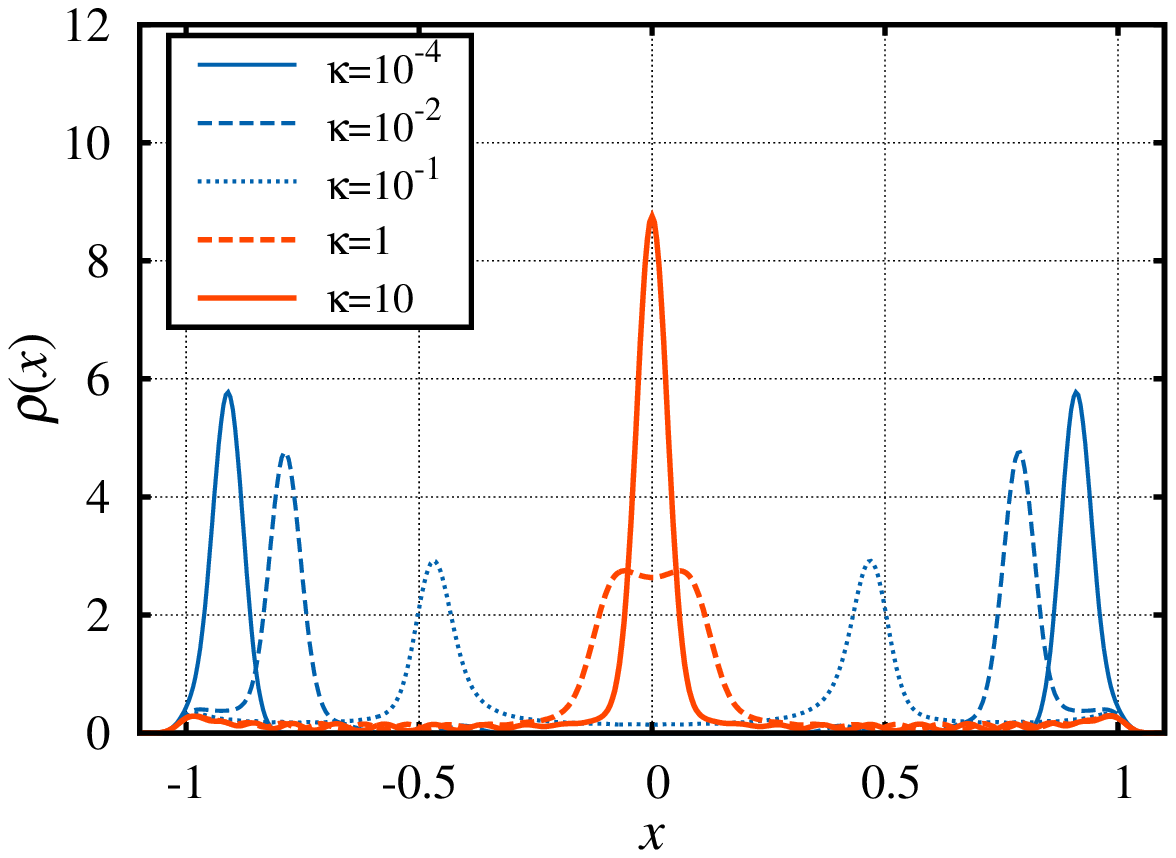}
\includegraphics[width=0.7\columnwidth]{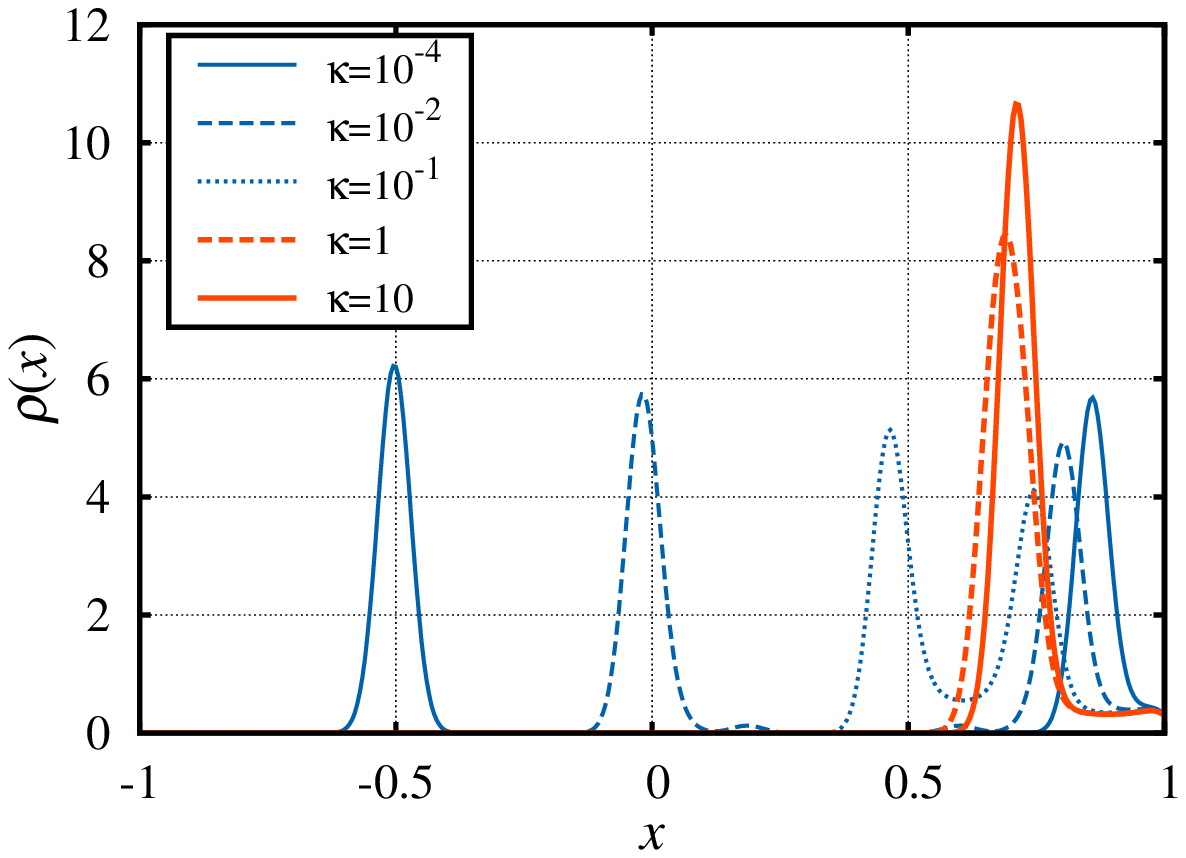}
\caption{Steady-state belief distribution under unbounded confidence and in the case of community-based influence, as the level of interaction
($\kappa$) varies. Stubborn users are present in both communities (top), or in only one community
(bottom).}
\label{fig:communities}
\end{figure}

\smallskip\noindent {\bf Community-based mutual influence.} 
Now, we consider a case of practical relevance where two communities
of users, with opposite biased prejudice, may interact. We express the
level of interaction between the two communities by the parameter
$\kappa$, and set the mutual influence to $\zeta(p,p')=
\frac{1}{2}+\frac{1}{2}\mbox{\rm erf}(p p'/\kappa)$.
Note that, the larger the $\kappa$, the stronger the mutual influence.
Fig.~\ref{fig:communities}(top) depicts the belief
distribution when the two communities exhibit a perfectly symmetric
structure in terms of agents' stubbornness and strength of
interactions within each community. The plot highlights the importance
of interaction between users belonging to two different communities,
in view of reaching a global agreement. Indeed, as $\kappa$ (hence the
interaction between communities) increases, users' beliefs tend to
mix, and an agreement is essentially reached for $\kappa=10$.

Fig.~\ref{fig:communities}(bottom) instead shows the belief
distribution in a similar scenario where stubborn users are
now present only in the  community characterized by biased positive
prejudice.
Interestingly, for very small values of $\kappa$ (i.e., when
the interaction within each community dominates), the community with
negative prejudice (the weak community) reaches a local agreement
around its center of mass, while the other (the strong community) remains 
anchored to the belief of its stubborn agents. As the
inter-community interaction increases, the beliefs of the weak
community tend to move toward those of the strong one. In particular,
for a given level of interaction,  
the distance between the communities' beliefs decreases significantly   
with respect to the case described in the top plot. We also remark 
that the average belief is now always biased toward the opinion of 
the strong community, further underlying the
importance of the role of stubborn users in belief dynamics. 

We remark that similar behaviors have been observed in sociology
studies such as~\cite{sociology-community} and references therein. 


\begin{figure}[tb]
  \centering
\includegraphics[width=0.7\columnwidth]{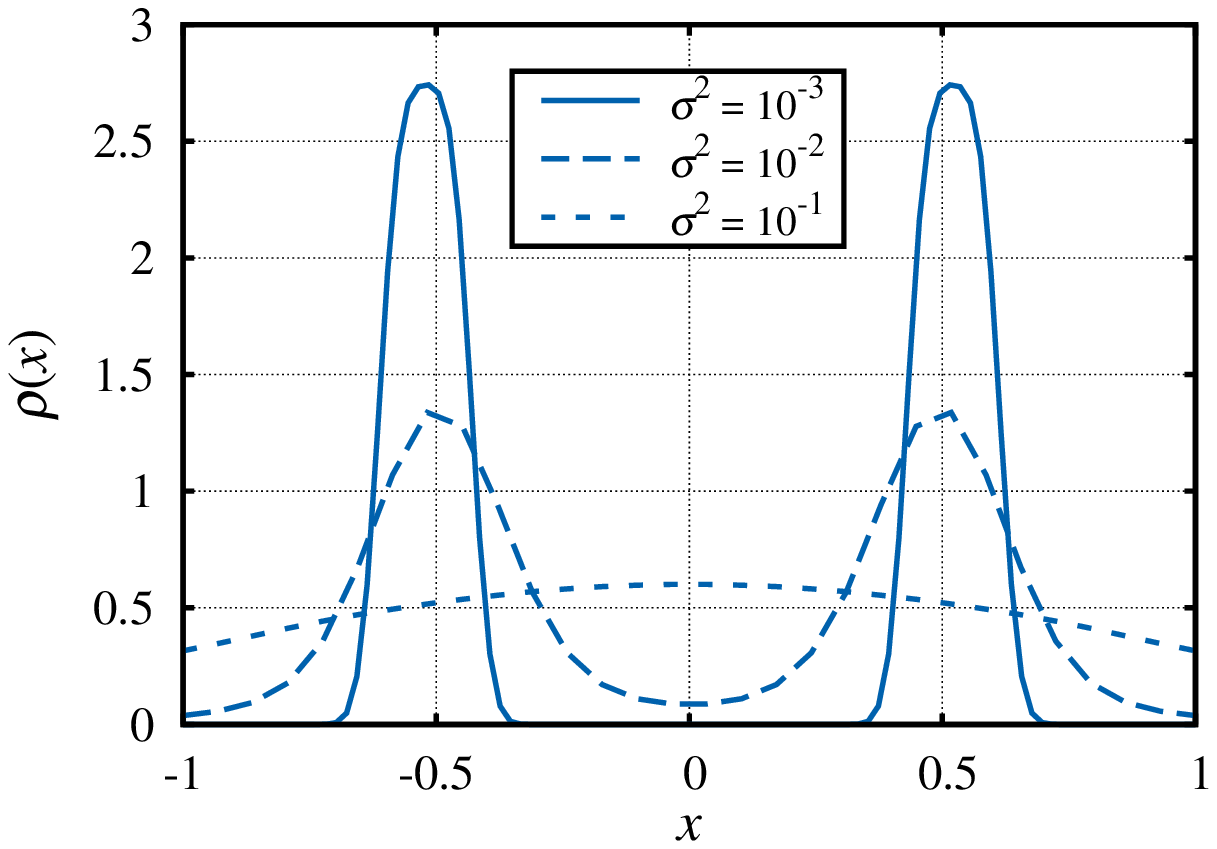}
\includegraphics[width=0.7\columnwidth]{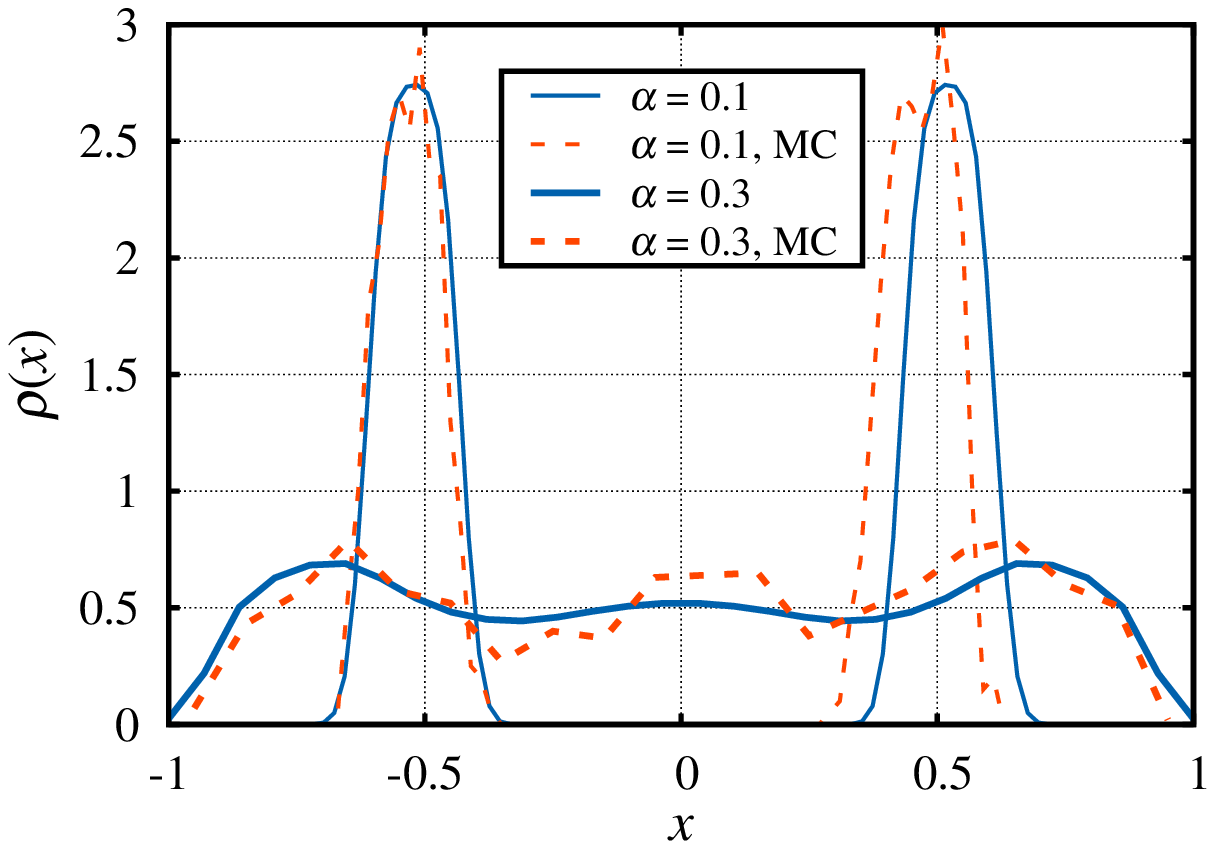}
\caption{\label{fig:bounded_12}Steady-state belief distribution under
  bounded confidence, when all agents have the same level of stubbornness and mutual influence vanishes 
with the distance between agents' beliefs: (top) impact of the endogenous noise; (bottom) impact of
  agents' stubbornness. The bottom plot also presents Monte Carlo (MC)
simulation results.}
\vspace{-3mm}
\end{figure}

\subsection{Bounded confidence\label{subsec:4.3}}
Here we consider the most challenging scenario in which bounded
confidence is accounted for. We recall that the results presented in
this case can be numerically obtained through the successive
approximation technique described in Sec.~\ref{sec:succapprox}.

We assume that the agents' personality is uniformly distributed in
$\Pc = [-1,1]$, agents' stubbornness is constant ($\alpha(p)=\alpha$),
and $u(p)=p$.  Also, the mutual influence is a smooth approximation of
a centered rectangular function~\footnote{$\zeta(|x-x'|, p, p')= \frac{1}{1+ (3|x-x'|)^{64}}$.}, whose
support is given by $|x-x'| \le 1/3$.  Fig.~\ref{fig:bounded_12}(top)
shows the belief distribution for $\alpha=0.1$ and different variance
of the endogenous noise.  As previously observed in the
literature~\cite{Garnier}, a clusterization of the agents' beliefs may
occur for effect of the bounded confidence.  In particular, for small
values of $\sigma^2$ (i.e., $\sigma^2=10^{-3}$) agents' beliefs are
well partitioned into two distinct clusters centered around $-0.5$ and
$0.5$, respectively.  The inter-distance of clusters is
sufficiently large that agents in the two clusters do not interact,
while agents within the same cluster are mutually attracted.  By
increasing $\sigma^2$, the spread of beliefs within each cluster
grows, and the clusters start to ``interfere'' ($\sigma^2=10^{-2}$) as
a consequence of the reduced distance between the cluster tails. Then
clusters completely disappear as the noise variance further
increases ($\sigma^2=0.1$).

Next, in Fig.~\ref{fig:bounded_12}(bottom) we look at the impact of agents' stubbornness. 
This parameter too plays a significant role: increasing $\alpha$ to 0.3 
makes clusterization vanish, even for small values of $\sigma^2$. The
reason is that the higher the $\alpha$, the weaker the mutual
attraction between agents' belief and, the smaller the deviation from
the original prejudice.  For the sake of validation of our
semi-analytical technique, Fig.~\ref{fig:bounded_12}(bottom) reports
also the long-run distribution of the agents' belief obtained with a
Monte Carlo simulator, in the case where the system includes $1000$
agents and beliefs evolve according to~\eqref{eq:x_i}.

\section{Transient analysis under unbounded confidence\label{sec:unbounded}}
This section addresses the time evolution of agent's beliefs under the
hypothesis of unbounded confidence (i.e., $\zeta(|x'-x|,p,p') = \zeta(p,p')$)
and $\Xc = \RR$.

In this case, the expression of $\mu_x(p,x,t,\rho)$ in \eqref{eq:mu_x2} becomes
\begin{equation} \label{eq:mu_x_trans}
\mu_x(p,x,t,\rho) 
= \bar{\alpha}(p) \eta(p) \left[ \phi(p, t) - x \right] + \alpha(p) [u(p)-x] \nonumber
\end{equation}
where $\eta(p)$ is defined in~\eqref{eq:etap} and 
\begin{equation} \label{eq:J1}
\phi(p,t) \triangleq \frac{1}{\eta(p)} \int_{\Zc} x' \zeta(p,p') \rho(p',x', t) \dd x' \dd p' ,
\end{equation}
is the extension of $\phi(p)$ (defined as the ratio of~\eqref{eq:betap} and~\eqref{eq:etap}) to the transient analysis.

In this case, the FP equation describes an Ornstein-Uhlenbeck random process,
whose Green function (impulse response) can be obtained by the method of
characteristics, as shown in Appendix E.
So doing, we obtain that, for a given value of $p$ and starting from a mass
point in $x=u(p)$, the agent density in the belief dimension is Gaussian with
mean
\begin{eqnarray}
m(p,t) & = &
\ee^{-w(p) t} u(p) + [1-\ee^{-w(p) t}] \frac{\alpha(p) u(p)}{w(p)} \nonumber\\
&& \quad+\bar{\alpha}(p) \eta(p)\int_0^{t} \ee^{w(p) (\tau-t)} \phi(p,\tau) \dd\tau 
\label{eq:mean_x}
\end{eqnarray}
and variance $\sigma^2(p,t) = \sigma^2(1-\ee^{-2w(p) t})/(2w(p))$
where, we recall that $w(p) = \alpha(p) +\bar{\alpha}(p) \eta(p)$.  We
remark that $m(p,t)$ in \eqref{eq:mean_x} is composed of three
terms. The first, which eventually fades away with time constant
$1/w(p)$, is due to the initial condition. The second term, which
includes a stationary contribution, contains the prejudice of the
agents with personality $p$. The third term is related to the
interaction between agents.  Also,
$\lim_{t \to \infty} m(p,t) = m(p)$, as defined in~\eqref{thetap}.

However, notice that $m(p,t)$ above is a function of $\phi(p,t)$,
which is in turn a function of $\rho(p,x,t)$ (see \eqref{eq:J1}).  As
such, we have to impose a self-consistency condition, as we have done
for the stationary analysis.  Precisely, the solution must satisfy
\begin{eqnarray}
\phi(p,t) 
&=&\frac{1}{\eta(p)} \int_{\Pc} \zeta(p,p')  \int_{\RR} x \rho(p',x,t) \dd x \dd p' \nonumber \\
&=&\frac{1}{\eta(p)}
\int_{\Pc} \zeta(p,p') \rho_0(p') \int_{\RR} x  
  \frac{\ee^{-\frac{(x-m(p',t))^2}{2 \sigma^2(p',t)}}}{\sqrt{2 \pi \sigma^2(p',t)}} \dd x \dd p' \nonumber \\
&=&\frac{1}{\eta(p)}\int_{\Pc} \zeta(p,p') \rho_0(p') m(p',t) \dd p' \nonumber  \\
&=&I_0(p,t) + I_1(p,t) +\frac{1}{\eta(p)}\int_{\Pc} \zeta(p,p')  \rho_0(p')
\bar{\alpha}(p')  \nonumber\\
&&\quad\cdot\eta(p') \int_0^{t} \ee^{w(p') (\tau-t)} \phi(p',\tau) \dd\tau \dd p' \label{eq:phi_pt} 
\end{eqnarray}
where we have defined for brevity:
\begin{equation}
I_0(p,t) \triangleq \frac1{\eta(p)} \int_{\Pc} \zeta(p,p')  \rho_0(p')
\ee^{-w(p') t} u(p')\dd p'  \nonumber
\end{equation}
\begin{equation}
I_1(p,t) \triangleq \!\frac1{\eta(p)} \int_{\Pc} \zeta(p,p')  \rho_0(p')
(1-\ee^{-w(p') t}) \frac{\alpha(p') u(p')}{w(p')}  \dd p' \,. \nonumber
\end{equation}
In order to solve~\eqref{eq:phi_pt} for $\phi(p,t)$, we take its Laplace transform over time
(whose variable will be denoted by $s$) and get
\begin{eqnarray} \label{eq:fixed_point_transform}
\widehat{\phi}(p, s) &=& \widehat{I}_0(p,s) \plus \widehat{I}_1(p,s)
 \plus \frac1{\eta(p)}\int_{\Pc}\zeta(p,p')\rho_0(p')\bar{\alpha}(p')  \nonumber\\
&& \quad\cdot \eta(p') \frac{\widehat{\phi}(p',s)}{s+ w(p')} \dd p' 
\end{eqnarray}
where $\widehat{\phi}(p,s)$, $\widehat{I}_0(p,s)$ and $\widehat{I}_1(p,s)$ are
the Laplace transforms of $\phi(p,t)$, $I_0(p,t)$ and $I_1(p,t)$, respectively.
Eq.~\eqref{eq:fixed_point_transform} is the integral equation for the transient
analysis, which corresponds to~\eqref{eq:beta-star2} in the stationary solution.

\begin{figure}[tb]
\centering
\includegraphics[width=0.48\columnwidth]{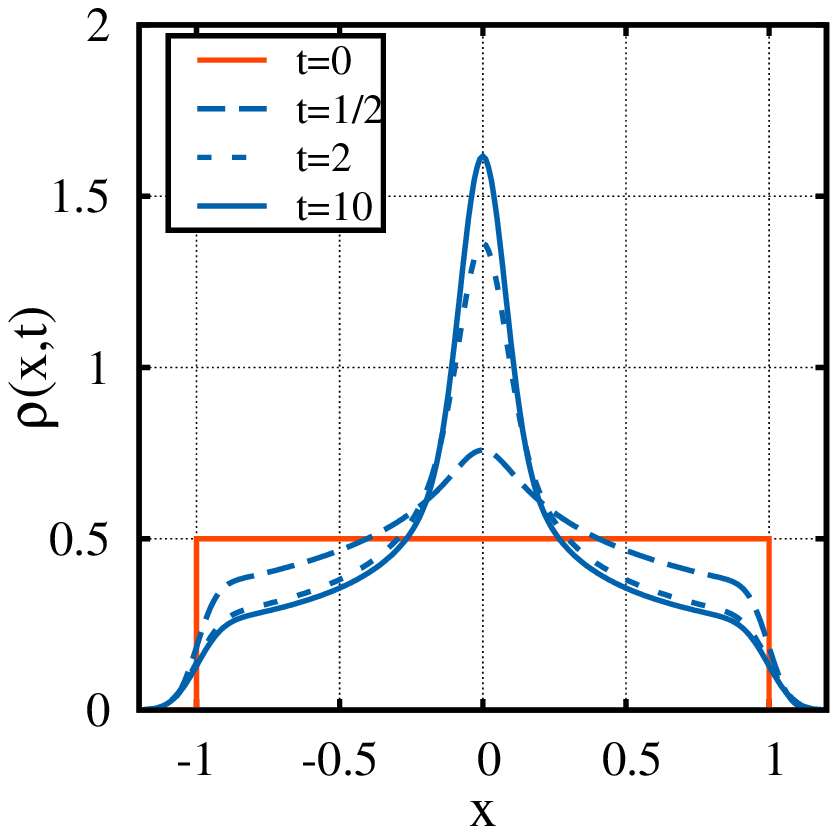}
\hfill
\includegraphics[width=0.48\columnwidth]{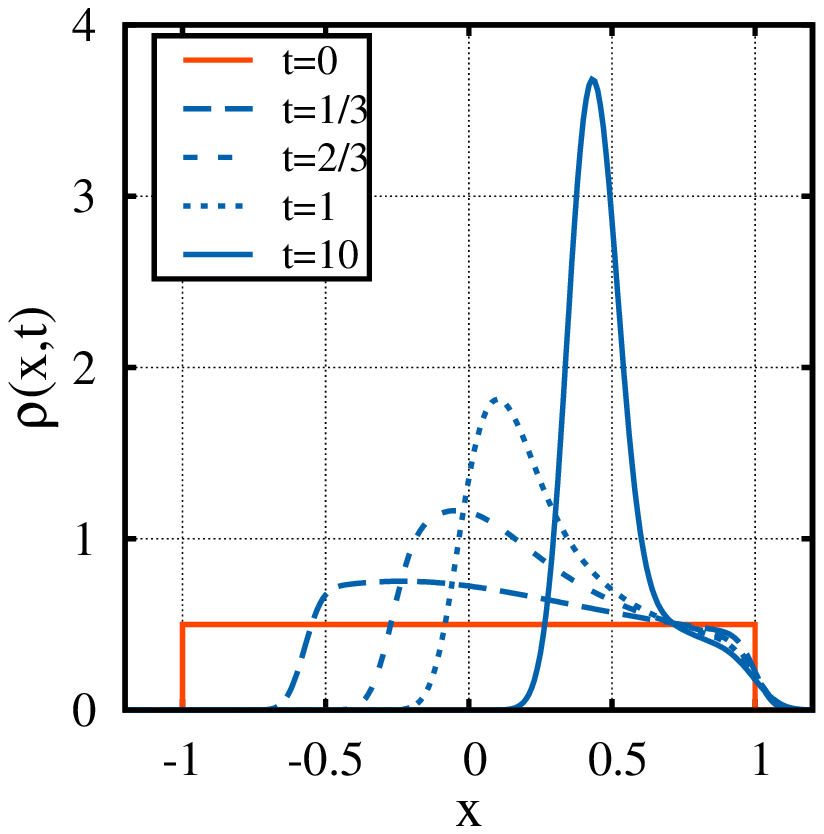}
\caption{Time evolution of the belief distribution for $\sigma^2 =0.01$ and constant mutual influence.
(Left) Stubborn users have biased prejudice ($\alpha(p) = |p|$); (Right) stubborn users have prejudice biased 
toward 1.
}
\label{fig:transient}
\vspace{-3mm}
\end{figure}

In the particular case where $\zeta(p,p') \equiv \zeta_1(p)\zeta_2(p')$,
we have that $\widehat{\phi}(p,s)$, $\widehat{I}_0(p,s)$ and $\widehat{I}_1(p,s)$
in~\eqref{eq:fixed_point_transform} do not depend anymore on $p$, and, as a
consequence, \eqref{eq:fixed_point_transform} reduces to
\begin{equation} \label{eq:J1_easy}
\widehat{\phi}(s) = \frac{\widehat{I}_0(s) + \widehat{I}_1(s)}{ 1 - \widehat{I}_2(s) } 
\end{equation}
where $\widehat{I}_2(s) = \int_{\Pc} \zeta_1(p)\zeta_2(p)\frac{\rho_0(p)\bar{\alpha}(p)}{s +
  w(p)}\dd p$.  In conclusion, we obtain an explicit solution for
$\widehat{\phi}(s)$ (in the Laplace domain), whose singularities are the
values of $s$ for which $\widehat{I}_2(s) = 1$.

Notice that, if $\alpha(p)$ takes only a finite number of values, so does $w(p)$
and the resulting $\widehat{\phi}(s)$ is a rational function, which can be
inverse-transformed. In particular, suppose $\alpha(p) \equiv \alpha$. Then
$w(p) \equiv w$ and all the integrals can be explicitly computed. If
$\overline{m}(t)$ is the average belief at time $t$, i.e.,
\begin{equation}
\overline{m}(t) \triangleq \int_{\Pc} m(p,t) \dd p\,,\nonumber
\end{equation}
it can be easily found that $\overline{m}(t) = \overline{u}$ where $\overline{u}$ is the average prejudice value.

\section{Results on the transient belief distribution\label{subsec:results-transient}}
\begin{figure}[tb]
  \centering
\includegraphics[width=0.7\columnwidth]{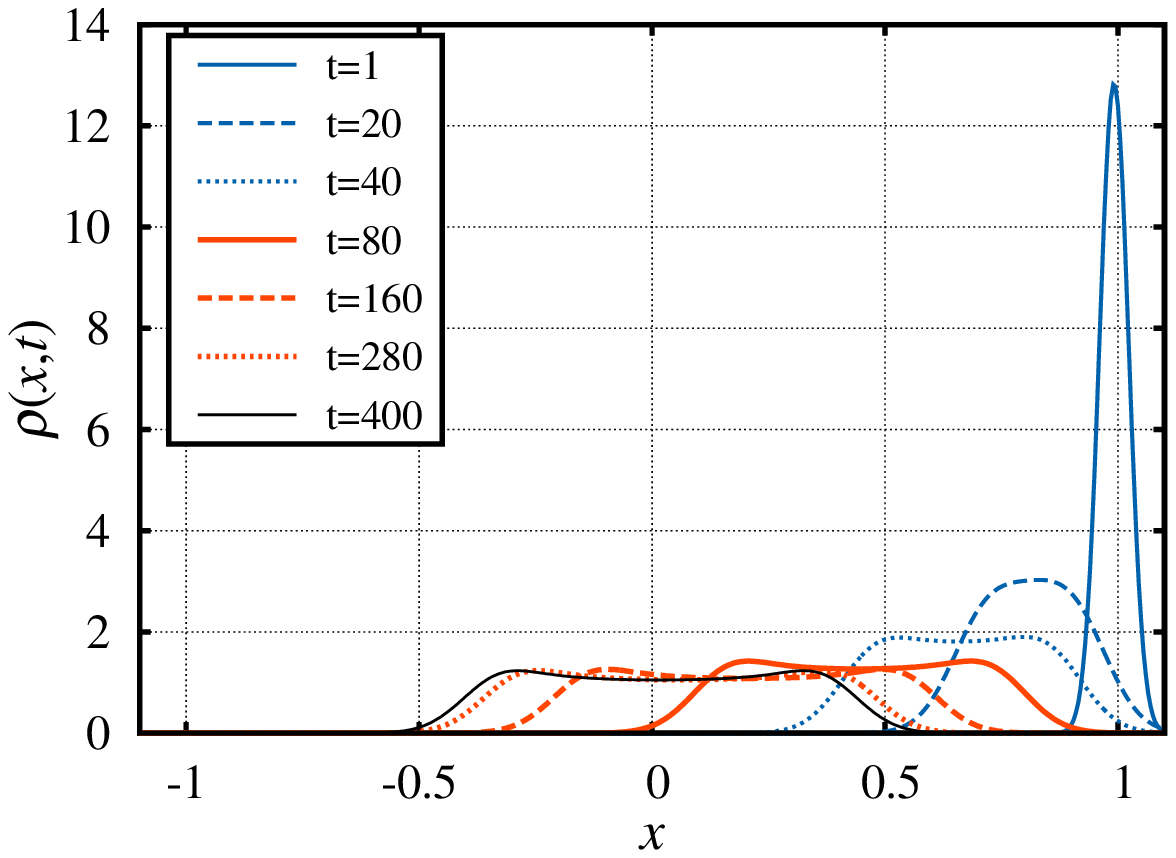}
\includegraphics[width=0.7\columnwidth]{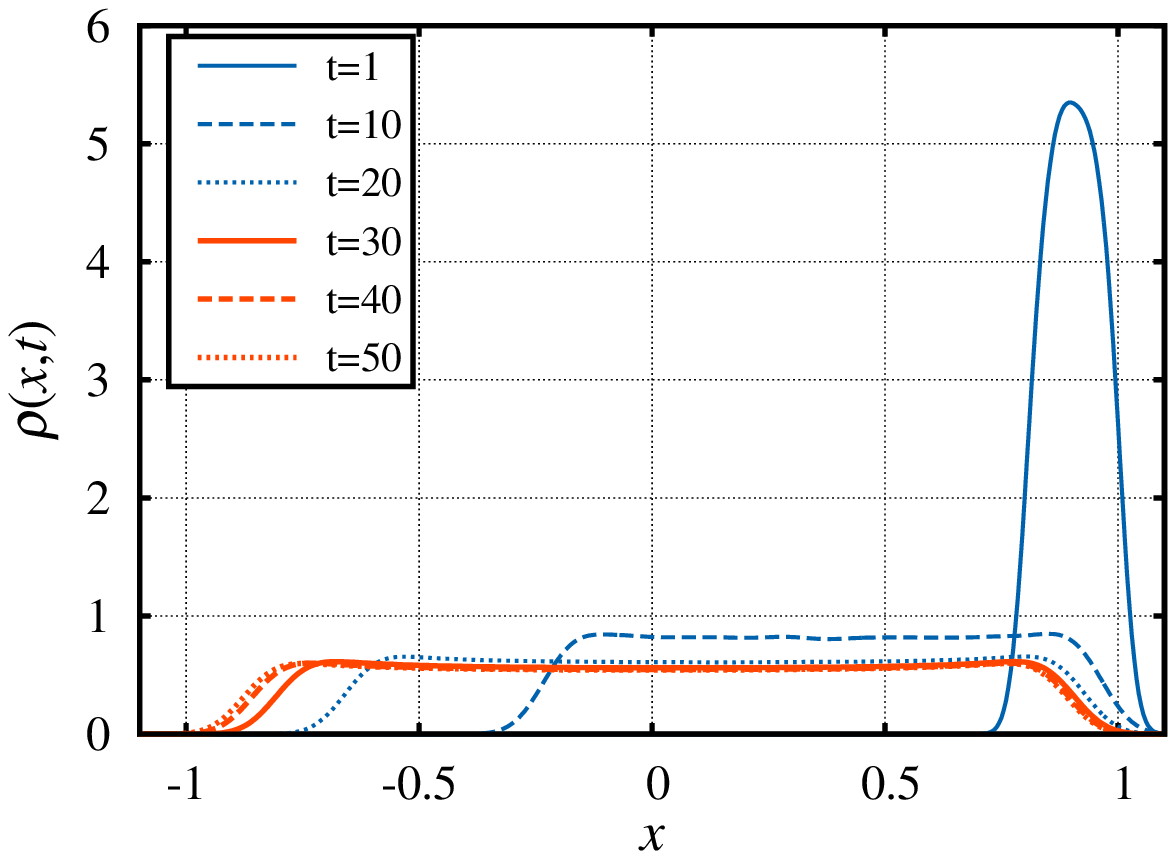}
\caption{Time evolution of the belief distribution when the mutual 
  influence increases with prejudice similarity and the initial belief
  density is Gaussian with mean equal to 1. The value of
  agents' stubbornness is constant: (top) low stubbornness
  ($\alpha=0.01$); (bottom) high stubbornness ($\alpha=0.1$).\label{fig:Transitorio_alpha_001-01}}
\vspace{-3mm}
\end{figure}

The time evolution of the belief distribution,
$\rho(x,t)=\int_{\Pc}\rho(p,x,t)\dd p$, is depicted in
Figs.~\ref{fig:transient} and~\ref{fig:Transitorio_alpha_001-01} for
$u(p)=p$.

\smallskip\noindent{\bf Constant mutual influence. }  Here we show the
results obtained for constant mutual influence (i.e.,
$\zeta=1$). Fig.~\ref{fig:transient}(left) refers to the case of
stubborn agents with extremal prejudices ($\alpha(p) =|p|$).
Fig.~\ref{fig:transient}(right), instead, corresponds to the case
where stubborn users are those with $p=1$  ($\alpha(p) = (p+1)^2/4$).
We remark that in both cases $\rho(p,x,t)$ can be computed through the
analytical expressions presented in the previous subsection.
Specifically, we have $\phi(p,t) = 0$ for the case outlined in
Fig.~\ref{fig:transient}(left), and
$\phi(p,t) = \frac{1}{2} \1_{\{t>0\}}$ for the case shown in
Fig.~\ref{fig:transient}(right). In both cases, the dominant time constant
is $1/w(p)=1$, and we have observed that $\rho(x,10)$ already closely
matches the steady-state distributions.

\smallskip\noindent{\bf Event-driven time evolution.}
Fig.~\ref{fig:Transitorio_alpha_001-01} shows the evolution of the
belief distribution over time, right after an event that has strongly
affected the agents' belief. We stress that such scenario has been
addressed in sociology studies, and similar phenomena have been
observed (see, e.g., \cite{sociology-dramatic}). A typical example is the impact of 
breaking news on the public opinion. We model such a situation with a
Gaussian distribution with mean equal to 1 and very small variance
($10^{-4}$).

We consider a case where agents show the same level of stubbornness,
and the closer the agents' prejudice, the stronger the interaction
($\zeta(p,p')=(1+(p-p')^2)^{-1}$).  The top plot refers to the
scenario with fickle agents ($\alpha=0.01$); in this case the
convergence to the steady-state distribution is slow.  A much faster
dynamic is instead observed in the bottom plot, which refers to a
higher level of user stubbornness ($\alpha=0.1$). In this case, after
the initial event, it takes just 1/10 of the time to return to the
steady-state distribution, suggesting that the time constant is
inversely proportional to the level of user stubbornness.

\section{Related work\label{sec:rel-work}}

Several models appeared in the literature represent social
interactions between agents through static graphs. 
In this  representation, often users directly interact only with
their neighbors, varying their beliefs for effect of pairwise ``attractive''
interactions~\cite{F&J,Tempo}.  In particular, agent $i$, interacting with agent
$j$, updates her belief, represented by a real number, to a new value, which is
a convex combination of her own original belief and the belief of agent $j$.
As an alternative, a growing thread of works considers highly dynamical
settings, in which every agent may interact, still through pairwise
``attractive'' interactions, with every other agent in the system
\cite{degroot1974reaching,Fagnani-Acemoglu}. Such models have been applied to
describe interactions through on-line forums like Reddit, Popurls or Newsvine,
as well as direct face-to-face interactions in crowded places such as meetings
and conferences.  A third thread of work has modeled social interactions through
a graph whose structure varies dynamically~\cite{shi2013agreement}. These pieces
of work capture the fact that in many social networks, e.g., Twitter, agents
dynamically follow or divert from other agents' beliefs based on
interests/beliefs similarity.

Another relevant class of models is represented by the so-called {\em
  bounded confidence}, in which social interactions occur only between
agents with similar beliefs
\cite{7-Deffuant,8-Hegselmann,Fagnani-Como,Garnier,9-Blondel,16-Weisbuch,17-Bhattacharyya,Varshney2014,Varshney2017}.
In particular, \cite{Varshney2014,Varshney2017} show how bounded
confidence models can be used to represent social interactions between
Bayesian decision makers.  Relevantly to our work,~\cite{Fagnani-Como}
models the social influence on the global opinion evolution in the
case of homogeneous agent interactions.  By adopting a mean-field
approach, the authors derive a differential equation whose solution
asymptotically (when the number of agents grows large) represents the
opinion evolution.  Mean-field theory is applied to the study of
opinion dynamics also in \cite{Garnier}, where Garnier et al. derive a
FP equation modeling the time evolution of beliefs in a neighborhood of
$t=0$, in the case of a large number of agents. Through
their model, the authors obtain conditions under which a
clusterization of the agents' beliefs occurs.  Another study that is
particularly relevant to our work is \cite{Baccelli}, where Baccelli
et al. propose a fairly general model that combines several features
of previous representations: (i) agent interactions constrained by a
graph, (ii) bounded confidence, and (iii) agent endogenous belief
random dynamics.  For such a model, the authors derive sufficient and
necessary conditions for stability, i.e., conditions under which the
relative spread of beliefs keeps finite.

We remark that our work significantly differs from the existing
studies. In particular, with respect to \cite{Fagnani-Como}, our
model, hence our analysis, is much more general, since it captures
non-homogeneous social interactions, the agents' prejudice and
personality, as well as exogenous random perturbations in the agents'
beliefs.  The effect of the agents' prejudice has been neglected also
in \cite{Garnier,Baccelli}. Furthermore, unlike \cite{Garnier}, which
focuses on early stages of the belief temporal evolution, we aim to
study the whole-time evolution of the system in the simple case with
no bounded confidence, and the steady-state ($t\to \infty$) under more
general conditions. Finally, we stress that the goal of our work is
not limited to obtaining results on the system stability (as in
\cite{Baccelli}), but it also studies the transient and steady-state
opinion dynamics.  

\section{Conclusions\label{sec:conclusions}}
Motivated by the great advent of social networks, we investigated the
evolution of beliefs within a social system and how they are affected
by the interactions between users. Our model captures all the main
features characterizing a social system. Through such model and a
mean-field approach, we analyzed the system behavior as the number of
users grows large.  In particular, by representing the belief
evolution through a Fokker-Plank partial differential equation, we
studied the steady state as well as the transient behavior of the
social system.  Our main results include: (i) closed-form expressions
for the steady-state distribution, which hold in some relevant cases,
and (ii) semi-analytical techniques to obtain the steady-state
distribution of social beliefs in the most general cases as well as
the transient distribution for a remarkably large class of systems.
Our analytical derivations are complemented with numerical results,
which show interesting dynamics due to social relationships and
different degrees of dependence on the users' prejudice.
 
Future work will address the model validation through
  experiments with real-world, Twitter data sets. To this end, by
  using existing APIs, we intend to collect tweets related to a 2-3
  weeks' time period around an event of particular interest, such as a
  political election. This set, organized in chronological order,
  should be filtered by using hashtags that well represent the
  selected event; then, a sentiment analysis tool should be run on the
  filtered set.  As done in our work, we can assume that the
  personality coincides with the prejudice, and we can take the first
  value of sentiment expressed by a user as her prejudice.
  Contextually, we will need to track the follower-followee
  relationship between users appearing in our trace and showing to be
  sufficiently active.  A fair (conservative) approximation would be
  to consider as follower a user that either re-tweets another user's
  message or write a new message referring to it.  Finally, we will
  use part of the obtained data as training set to estimate the model
  parameters, and the rest as testing to validate the prediction
  capabilities of our model.

\bibliographystyle{IEEEtran}
\bibliography{Refs}
\appendices
\newpage
\onecolumn

\section{Ergodicity of the dynamic system described by (1)}\label{app:drift}
We consider Equation (1) in the main manuscript,  with a finite number
of agents ($U$) and we show that the agents' belief exhibits a
negative drift outside a compact domain,  therefore  the associated Markov chain is ergodic.

First, it can be easily seen  that ours is not an  explosive Markovian
model (see
\cite[Sec. 20.3.1]{Tweedie-Meyn}). 
Second, we remark that a time-sampled version of our model can be seen
as a special case of  the Non-Linear State Space (NSS(F))
model defined in \cite{Tweedie-Meyn}. 
Thus, in light of the  fact that  the noise term in (1) is Gaussian
and additive, the Markov chain describing the dynamics of agents's
belief  
results to be:

i)   $\psi$-irreducible over $\Zc=\Xc\times \Pc$,  
where measure $\psi$ is equal to the product measure generated by the
Lebesque measure  over $\Xc=\mathbb{R}$ 
and a discrete measure on $\Pc$  
(see again \cite[Sec. 20.3.1 and Th. 6.0.1]{Tweedie-Meyn});

ii) forward accessible  and, hence,   a $T$-chain (see  \cite[Prop. 7.1.5]{Tweedie-Meyn}).

Now observe that we can rewrite (1)  as:
\[  X_i(t+\dd t) = X_i(t)  + \Delta X_i(t)dt  + \sigma d W_i \]
where the belief drift $\Delta X_i(t)= \frac{ [1-\alpha(P_i)]}{U}\sum_{j\in \Uc, j \neq i} \zeta(|X_i(t)-X_j(t)|,P_i,P_j)(X_j(t)-X_i(t))   + 
\alpha(P_i) (u(P_i)-X_i(t))$. 

Next, for ease of presentation, we limit ourselves to consider the set of agents with
positive belief $X_i(t)$, the same derivation applies to negative
values of $X_i(t)$ (just changing the sign of both belief and drift).
If we consider  an agent  $k$ s.t. $k\in \argmax_i X_i(t)$, then:
\begin{eqnarray*}
 \Delta X_k(t)= 
&=&   \frac{ [1-\alpha(P_k)]}{U}\sum_{j\in \Uc, j \neq k}   \zeta(|X_k(t)-X_j(t)|,P_k,P_j)(X_j(t)-X_k(t))  + 
\alpha(P_k) (u(P_k)-X_k(t))  \\
&\le& [\alpha(P_k) u(P_k)-  \alpha(P_k) X_k(t)]\\
&\le& [\sup_k \alpha(P_k) u(P_k) - \inf_k
\alpha(P_k)X_k(t)] 
\end{eqnarray*}
where we bounded the first term on the RHS with zero since, by construction, $X_j(t)-X_k(t)=X_j(t)-\max_i X_i(t)\le 0$.

Similarly, if we consider an agent $l$,  we have:
\begin{eqnarray*}
 \Delta X_l(t)
&=&   \frac{ [1-\alpha(P_l)]}{U}\sum_{j\in \Uc, j \neq l}
 \zeta(|X_l(t)-X_j(t)|,P_l,P_j)(X_j(t)-X_l(t)) 
 + 
\alpha(P_l) (u(P_l)-X_l(t)) \\
&=&   \frac{ [1-\alpha(P_l)]}{U}\sum_{j\in \Uc, j \neq l}
 \zeta(|X_l(t)-X_j(t)|,P_l,P_j)
[(X_j(t)-X_k(t))+(X_k(t)-X_l(t))]    \\
&+ & \alpha(P_l) (u(P_l)-X_l(t))  \\
&\le& [\sup_i \alpha(P_i)  u(P_i) -  \alpha(P_l) X_l(t))+ (1-\alpha(P_l)) S_\zeta  (X_k(t)-X_l(t))] 
\end{eqnarray*}
where   $S_\zeta =
 \sup_{x_i,x_j,P_i,P_j} \zeta(|x_i-x_j|,P_i,P_j)<\infty$ with $x_i,
 x_j \in \mathbb{R}$. 
By considering sufficiently large $X_l(t)$ such that $X_l(t)=(1- \beta_l) X_k(t)$ (with $\beta_l<1$), we obtain: 
 \[
 \Delta X_l(t)\le [\sup_i \alpha(P_i) u(P_i) -  \alpha(P_l)(1-
 \beta_l) X_k(t)+ (1-\alpha(P_l))   S_\zeta  \beta_l X_k(t)]\le 
 \sup_i \alpha(P_i) u(P_i) - \epsilon X_k(t)
 \]
for any arbitrarily chosen $0	<\epsilon < \inf_i\alpha(P_i)$ 
whenever $\beta_l \leq\frac{\alpha(P_l)-\epsilon}{\alpha(P_l)+ S_\zeta (1- \alpha(P_l))}$. 
Thus,  given $\beta_0=\inf_i \frac{\alpha(P_i)-\epsilon}{\alpha(P_i)+
  S_\zeta (1- \alpha(P_i))}>0$, we have that 
 $ \Delta X_l(t)\le\sup_i \alpha(P_i)  u(P_i) - \epsilon X_k(t) $,
 uniformly  over all $l$ such that $X_l(t)> (1- \beta_0)X_k(t)$.
 At last, observe that the following holds  uniformly  on every $l$
 whenever $ \sup_\alpha(P_i)  u(P_i)\le    \alpha(P_k) S_\zeta X_k(t)$:
 \[
 \Delta X_l(t)\le   [\sup_i \alpha(P_i) u(P_i) +
 (1-\alpha(P_l)  S_\zeta  X_k(t)]\le  
 S_\zeta  X_k(t) \,.
 \]

Next, let us define the following Lyapunov function: 
$\mathcal{L}(t)= \frac{1}{n+1}\sum_i X_i(t)^{n+1}$  for a
sufficiently large odd $n$ such that 
 $\epsilon/3 X_k(t)^{n+1}>U  S_\zeta  X_k(t)\left [(1-
   \beta_0)X_k(t)\right]^n$, hence  
 $\epsilon/3 >U  S_\zeta  (1- \beta_0)^n$.

 Then, we can write:
 \begin{eqnarray*}
 \Delta \mathcal{L}(t)&=& \mathbb{E}\left [ \mathcal{L}(t+ \dd t)-
\mathcal{L}(t)\mid \{X_i(t)\}_i \right ] \\
&=& \mathbb{E}\left [ \sum_i X_i^n(t)\cdot ( \Delta X_i(t)\dd t  + \sigma \dd W_i) \right ] 
+o \left( \mathbb{E} \left [\sum_i X_i^n(t)\cdot ( \Delta X_i(t) \dd t
    +\sigma \dd W_i )\right ] \right) \,.
\end{eqnarray*}

Now, let us focus on the first term on the RHS of the previous
expression and consider that $X_k(t)$ is sufficiently large. We get: 
\begin{eqnarray*}
 \mathbb{E}\left [\sum_i X_i^n(t) \cdot ( \Delta X_i(t)\dd t  +
   \sigma \dd W_i) \right ] \nonumber\\
&=& \sum_i X_i^n(t)\cdot  \Delta X_i(t)\dd t \label{lyapu1}\\ 
&=&\sum_{l: X_l(t)>(1-\beta_0)X_k(t)} X_l^n(t)\cdot \Delta X_l(t)\dd
t  \nonumber\\
&+ &\sum_{l:X_l(t)\le (1-\beta_0)X_k(t)} X_l^n(t)\cdot \Delta X_l(t)\dd t \nonumber\\
&\le&  X_k^n(t) [\alpha(P_k) u(P_k)-  \alpha(P_k) X_k(t)]\dd t \nonumber\\
&+ &\sum_{l\neq k: X_l(t)>(1-\beta_0)X_k(t)} X_l^n(t)\cdot [\sup_i \alpha(P_i) u(P_i) - \epsilon X_k(t)] \dd t \nonumber\\
&+& \sum_{l: X_l(t)\le (1-\beta_0)X_k(t)} X_l^n(t)\cdot  S_\zeta  X_k(t) \dd t\nonumber\\
&\le&  X_k^n(t) [\sup_i\alpha(P_i) u(P_i)-  \inf_i \alpha(P_i)
X_k(t)]\dd t + \frac{\epsilon}{3} X_k(t)^{n+1} \dd t \,.  \label{lyapu2}
\end{eqnarray*}
Note that we obtained the second inequality from the first one by
bounding the second sum with zero, which holds true when $X_k$ is
sufficiently large.

Now observe that  the term 
$o\left(\mathbb{E} \left[\sum_i X_i^n(t)\cdot ( \Delta X_i(t)\dd t  +
    \sigma \dd t W_i) \right ] \right)$, 
 can be assumed smaller than 
$\epsilon/3 X_k(t)^{n+1} \dd t$, for $X_k(t)$
sufficiently large. Thus, we have:
\begin{eqnarray*}
 \Delta \mathcal{L}(t) &\le &
 X_k^n(t) \sup_i\alpha(P_i) u(P_i)\dd t-  \inf_i \alpha(P_i)
 X_k^{n+1}(t)\dd t +
\frac{2\epsilon}{3} X_k^{n+1}(t) \dd t\\
& \le & X_k^n(t) \sup_i\alpha(P_i) u(P_i)\dd t- \frac{\epsilon}{3}
X_k^{n+1}(t) \dd t 
\end{eqnarray*}
where we exploited the fact that $\epsilon < \inf_i \alpha(P_i)$.
At last, chosing $B$  such that
\[
X_k(t)^n\sup_i\alpha(P_i) u(P_i)\dd t \leq \max\left( B^n
  \sup_i\alpha(P_i) u(P_i)\dd t, \frac{\epsilon}{12} X_k(t)^{n+1} \dd
  t \right ),
  \] 
we can bound:
\begin{eqnarray*}
 \Delta \mathcal{L}(t) &\le & 
   B^n \sup_i\alpha(P_i) u(P_i)\dd t- \frac{\epsilon}{4} X_k^{n+1}(t) \dd t \,.\\
\end{eqnarray*}
It follows that  the drift  of $\mathcal{L}(t)$ becomes negative
outside a compact set.  
Now,  in light of the fact that our system is a $\psi$-irreducible
$T$-chain,  
every compact set is petite (see \cite[Th. 6.2.5]{Tweedie-Meyn}). Consequently, we can invoke the counterpart   
of  \cite[Th. 14.2.3]{Tweedie-Meyn} for continuous-time processes, in order 
 to state that our Markov chain is $\max|X_l|^{n+1}$-regular, and therefore ergodic.


\section{Continuity of the map $\phi(p)\to\rho_{\phi(p)}(p,x)$}\label{app:map-continuity}

Our goal is to show that the mapping $\phi(p) \to \rho_{\phi(p)}(p,x)$
is continuous under $\mathbb{L}_\infty$ norm, i.e., that
\[
\| \rho_{\phi(p)}(p,x)- \rho_{\phi^*(p)}(p,x)\|_\infty\to 0
\] 
whenever
$\phi(p)\to \phi^*(p)$ uniformly, for any arbitrary $\phi^*(p)$.

We recall that $w(p) = \alpha(p) + \bar{\alpha}(p)\eta(p)$. By dropping the dependency on $p$ in the following expressions, we have
\begin{eqnarray}
  & & \hspace{-6ex}\| \rho_{\phi}(x)- \rho^*_{\phi}(x)\|_\infty\nonumber \\
  &=& \left\|\sqrt{\frac{w}{\pi\sigma^2}} 
      \left(\ee^{ -\frac{w[x- m]^2}{\sigma^2}}-\ee^{-\frac{w[x- m^*]^2}{\sigma^2}}\right) \rho_0\right\|_\infty \nonumber \\
  &\le& K^\infty \left\|\ee^{ -\frac{w[x- m]^2}{\sigma^2}}-\ee^{-\frac{w[x- m^*]^2}{\sigma^2}}\right\|_\infty\label{eq:cont1a} 
\end{eqnarray}
where $K^\infty=\|\sqrt{\frac{w}{\pi\sigma^2}} \rho_0 \|_\infty$,
$m$ is defined in~\cite[eq.~(15)]{OurPaper} and $m^*$ is given in~\cite[eq.~(15)]{OurPaper}
where $\phi$ is replaced with $\phi^*$. Observe that,
for any continuously differentiable function $f(x)$, we can write
$f(x+m-m^*) = f(x) + \int_0^{m-m^*} f'(x+y)\dd y = f(x) + (m-m^*)
  f'(x+\xi)$ with $\xi \in[0,m-m^*]$,\footnote{Without lack of generality we assume $m>m^*$.} where in the last equality we used
the mean value theorem for integrals. Thus, 
$|f(x+m-m^*)-f(x)| \le |m-m^*|\sup |f'(x)|$.  Then we have
\begin{eqnarray}
  \left\| \ee^{-\frac{w(x-m)^2}{\sigma^2}}\mathord{-}\ee^{-\frac{w(x-m^*)^2}{\sigma^2}} \right\|_\infty
  &\le& |m\mathord{-}m^*|\sup_x \left[\frac{\dd  e^{-\frac{w x^2}{\sigma^2}}}{\dd x}\right] \nonumber\\
  &=& |m\mathord{-}m^*|\sup_x \frac{2wx}{\sigma^2}\ee^{-\frac{wx^2}{\sigma^2}} \nonumber\\
  &<& \sqrt{\frac{2w}{\sigma^2 \ee }}|m-m^*|\,.\label{eq:cont2a}
\end{eqnarray}
Therefore, combining~\eqref{eq:cont1a} and~\eqref{eq:cont2a}, we have:
\begin{align*}
&\left|  
\ee^{ -\frac{w(x- m)^2}{\sigma^2}}-
\ee^{ -\frac{w(x-m)^2}{\sigma^2}}\right| \le
\sqrt{\frac{2w}{\sigma^2 \ee }}\frac{\bar{\alpha}\eta}{w}|\phi-\phi^*|
\end{align*}
and finally:
\[
\| \rho_{\phi}(x)- \rho_{\phi^*}(x)\|_\infty  
\le K^\infty\sqrt{\frac{2w}{\sigma^2 \ee }} \frac{\bar{\alpha}\eta}{w}|\phi^*-\phi| 
\]
which goes uniformly to $0$ as $\phi \to \phi^*$ uniformly, under the assumption that $\sup_p\rho_0(p)<\infty$.


\section{Continuity of the operator $\Ac$}\label{app:continuity}
\begin{teorema}
 The operator $\Ac$  is continuous  at the fixed point with respect to
 norm $\mathbb{L}_1$ convergence,  whenever $\Xc$  compact, or 
$\Xc=\mathbb{R}$ and $\zeta(p,p',|x-x'|)=0$  for $|x-x'|>X_0<\infty$.
\end{teorema}

We now prove the continuity of the operator $\Ac\{\cdot\}$ around the fixed point.
To this end, let $\rho^*(p,x)$ be the fixed point of the operator $\Ac\{\cdot\}$,
i.e., $\rho^*(p,x) = \Ac\{\rho^*(p,x)\}$. Let us consider a perturbation of the fixed point solution  
\[  \tilde{\rho}(p,x) = \rho^*(p,x) + \epsilon\rho(p,x)\]
where, without lack of generality we assume $\|\rho(p,x)\|_1=1$. 
We have to show that 
\[ \lim_{\epsilon\to 0}\Ac\{\tilde{\rho}(p,x)\} = \Ac\{\rho^*(p,x)\}=\rho^*(p,x)\]
To this purpose we define
\begin{equation}\label{eq:mux1}
\mu_{x,1}(p,x,\rho)=  \bar{\alpha}(p)\int_{\Zc}\zeta(|x'-x|) (x'-x)  \rho(p',x')\dd x'\dd p'\,.
\end{equation}
In the following we drop the dependency on $p$ when not necessary.
We note that
\[ \mu_{x,1}(x,\tilde{\rho})-\mu_{x,1}(x,\rho^*) = \mu_{x,1}(x,\tilde{\rho}-\rho^*) \]
since $\mu_{x,1}(x,\rho)$ is linear w.r.t. the parameter $\rho$.
Therefore, we can write the following bound:
\begin{eqnarray}
|\mu_{x,1}(x,\tilde{\rho}-\rho^*)|
&=&|\mu_{x,1}(x,\epsilon\rho(x))|  \nonumber \\
&\le& \bar{\alpha}\epsilon S_\zeta X_0  \int_{\Zc} \rho(p',x')  \dd x' \dd p'  \nonumber \\
& \le& \epsilon S_\zeta  X_0 
\end{eqnarray}
where we exploited the fact that $\rho(x)$ is a distribution,
$S_\zeta = \sup_{|x'-x|,p,p'} \zeta(|x'-x|,p,p')<\infty$ and we assumed
that $\zeta(|x'-x|,p,p')=0$ for $|x'-x|>X_0$. Therefore, using
\cite[eq.~(10)]{OurPaper}, we can write:
\begin{eqnarray} \label{eq:fixed-point-cont}
  \Ac\{\tilde{\rho}(x)\} 
  &\hspace{-2ex}= &\hspace{-2ex} K\ee^{\frac{2\bar{\alpha}}{\sigma^2}\int_{0}^x \int_{\Pc, x'}\zeta(|x'-y|) (x'-y) \rho^*(p',x') \dd p' \dd x'\dd y} \nonumber \\
  &\hspace{-2ex}  &\hspace{-1ex} \cdot \ee^{\frac{2\bar{\alpha}}{\sigma^2}\int_{0}^x \int_{\Pc,x'}\zeta(|x'-y|) (x'-y) \epsilon \rho(p',x') \dd p' \dd x'\dd y} \nonumber \\
  &\hspace{-2ex}  &\hspace{-1ex} \cdot\ee^{-\frac{\alpha[x-u]^2}{\sigma^2}} \rho_0 \nonumber\\     
  &\hspace{-2ex}= &\hspace{-2ex} K\ee^{\frac{2}{\sigma^2} \int_{0}^x \mu_{x,1}(y,\rho^*)\dd y}\ee^{\frac{2}{\sigma^2} \int_{0}^x \mu_{x,1}(y,\epsilon\rho)] \dd y}\nonumber \\
  &\hspace{-2ex}  &\hspace{-1ex} \cdot\ee^{-\frac{\alpha [x-u]^2}{\sigma^2}} \rho_0
\end{eqnarray}
Now, defined $v(x)=\frac{1}{x}\int_{0}^x\mu_{x,1}(y,\epsilon\rho)\dd y$  (with $v(0)=0$), 
we have:
\begin{eqnarray}\label{eq:Arho}
  \Ac\{\tilde{\rho}(x)\}
  &=& K' \ee^{\frac{2}{\sigma^2}\int_{0}^x  \mu_{x,1}(y,\rho^*) \dd y }\cdot \ee^{-\frac{\alpha}{\sigma^2}[x-u-v(x)]^2}\rho_0  \nonumber \\
\end{eqnarray}
where $K'=K \ee^{-[2 u v(x)+v(x)^2/\alpha]}$.
Since  $|\int_{0}^x\mu(y,\epsilon\rho) \dd y|\le \epsilon S_\zeta X_0 x$, and
\[ \Ac\{\rho^*(x)\}=\rho^*(x) = K \ee^{2\frac{ \int_{0}^x \mu_{x,1}(y,\rho^*) \dd y}{\sigma^2}}   
      \ee^{-\frac{\alpha (x-u)^2}{\sigma^2}}  
 \rho_0 \]
subtracting $\Ac\{\rho^*(x)\}$ from both sides of~\eqref{eq:Arho}, we have:
\begin{eqnarray}\label{eq:h+-}
  && \hspace{-8ex}\Ac\{\tilde{\rho}(x)\} -\Ac\{\rho^*(x)\}\nonumber \\
  &=&  K'\ee^{2\frac{\int_{0}^x  \mu_{x,1}(y,\rho^*) \dd y }{\sigma^2}}  
 \ee^{-\frac{\alpha [(x-u- v(x))^2}{\sigma^2}} \rho_0\nonumber \\
&&- K \ee^{2\frac{ \int_{0}^x \mu_{x,1}(y,\rho^*) \dd y}{\sigma^2}}   
      \ee^{-\frac{\alpha [(x-u)^2}{\sigma^2}}  
   \rho_0 \nonumber \\
&\le&  \rho^*(x)\left(\frac{K'}{K}\ee^{-\frac{\alpha [x-u- v(x)]^2}{\sigma^2}+\frac{\alpha [(x-u)^2}{\sigma^2}} -1\right)\,.
\end{eqnarray}
Note that
\[ \ee^{-\frac{\alpha [x-u- v(x)]^2}{\sigma^2}+\frac{\alpha [(x-u)^2}{\sigma^2}} = \ee^{-\frac{\alpha v(x)^2}{\sigma^2}+\frac{2\alpha(x-u)v(x)}{\sigma^2}}\]
Since $|v(x)|\le \epsilon S_\zeta X_0\to 0$ as $\epsilon\to 0$, then the above quantity tends to 1
uniformly. Similarly,
$K'/K \to 1$ uniformly for any $p$. Thus, the term $|\Ac\{\tilde{\rho}(x)\} -\Ac\{\rho^*(x)\}| \to 0$ as
$\epsilon\to 0$.

When $\Xc$  is a finite interval,  continuity of the operator $\Ac\{\cdot\}$ can
be proven under general conditions  on $\zeta(p,p',|x-x'|)$ by   uniformly bounding  $|\mu_{x,1}(x,\tilde{\rho}-\rho^*)|$ and then 
applying the same arguments as before. 


\section{On the local/global contraction properties of operator $\Ac\{\cdot\}$
  under  compact $\Xc$}\label{app:contraction}

Without loss of generality, let us consider the set $\Xc$ as symmetric with respect to 0 and define
\[\|\rho(p,x)\|_1= \int_{\Zc} |\rho(p,x)| \dd x  \dd p\]
Observe that the class of probability density functions (i.e., functions
$\rho(p,x)\ge 0$ with $\|\rho(p,x)\|_1= 1$) forms an invariant set under the
operator $\Ac\{\cdot\}$ (i.e., any probability density function $\rho(p,x)$ is mapped by
$\Ac$ onto a probability density function).  Furthermore, note that if
$\rho^{(k)}(p,x)$ is a probability density, then
$\rho^{(k+1)}(p,x)=\Ac\{\rho^{(k)}(p,x)\}$ belongs to the same class. It
follows that we can limit our analysis to the case where the operator $\Ac\{\cdot\}$ is
applied on probability densities.

In order to show that the operator $\Ac\{\cdot\}$ is a contraction, we need to show that
for any two probability densities, $\rho_1(p,x)$ and  $\rho_2(p,x)$, we have~\cite{kolmogorov2012introductory}: 
\begin{equation}
\| \Ac\{\rho_2(p,x)\}-\Ac\{\rho_1(p,x)\}\|_1 < c\| \rho_2(p,x)-\rho_1(p,x)\|_1
\label{eq:contraction}
\end{equation}
where $c<1$ and where $\|\cdot\|_1$ is the above defined norm.
We first apply the operator $\Ac\{\cdot\}$ on $\rho_1(p,x)$ and $\rho_2(p,x)$
separately. By the definition of $\Ac\{\cdot\}$, we have
\[ \Ac\{\rho_1(p,x)\} =  K_1(p)  \ee^{\frac{2}{\sigma^2}\int_0^x \mu_{x,1}(p,y, \rho_1)\dd y} 
\ee^{-\frac{\alpha(p) [x-u(p)]^2}{\sigma^2}} \rho_0(p) \]
where the positive normalization factor $K_1(p)$ is such that
\[ K_1(p)^{-1} = \int_\Xc \ee^{\frac{2}{\sigma^2}\int_0^x
  \mu_{x,1}(p,y, \rho_1) \dd y}
\ee^{-\frac{\alpha(p)[x-u(p)]^2}{\sigma^2}} \dd x \,.\]

To proceed further, we observe that $\mu_x(p,y, \rho)$ is linear in
$\rho$. Indeed~\cite[eq.~(4)]{OurPaper}, we have
$\mu_{x,1}(p,y, \rho_1+\rho_2) = \mu_{x,1}(p,y, \rho_1)+
\mu_{x,1}(p,y, \rho_2)$.  Let us define
$M_{x,1}(p,\rho)=\frac{2}{\sigma^2}\int_0^x \mu_{x,1}(p,y, \rho) \dd
y$ and
$G(x,p) = \ee^{-\frac{\alpha(p) [x-u(p)]^2}{\sigma^2}}\rho_0(p)$. It
follows
\begin{eqnarray}
\Ac\{\rho_2(p,x)\} 
&=& K_2(p) \ee^{M_{x,1}(p,\rho_2)}\ee^{-\frac{\alpha(p) [x-u(p)]^2}{\sigma^2}} \rho_0(p)\nonumber \\
&=& K_2(p) \ee^{M_{x,1}(p,\rho_1+(\rho_2-\rho_1))} G(x,p)\nonumber \\
&=& K_2(p) \ee^{M_{x,1}(p,\rho_1)+M_{x,1}(p,\rho_2-\rho_1)}G(x,p)\nonumber \\
&=& K_2(p) \ee^{M_{x,1}(p,\rho_1)}\ee^{M_{x,1}(p,\rho_2-\rho_1)}G(x,p)\nonumber \\
&=& \Ac\{\rho_1(p,x)\}\frac{K_2(p)}{K_1(p)}\ee^{M_{x,1}(p, \rho_2-\rho_1)}
\end{eqnarray}
with 
\begin{equation} 
\label{eq:K2}
K_2(p)^{-1} = \int_\Xc \ee^{M_{x,1}(p,\rho_1)}
\ee^{M_{x,1}(p,\rho_2-\rho_1) }\ee^{-\frac{\alpha(p)
    [x-u(p)]^2}{\sigma^2}} \dd x \,.
\end{equation} 
Therefore, 
\begin{eqnarray}
& &\hspace{-4ex}\|\Ac\{\rho_2(p,x)\}-\Ac\{\rho_1(p,x)\}\|_1 \nonumber \\
&\le& \left\|\Ac\{\rho_1(p,x)\}\left(  \frac{K_2(p)}{K_1(p)} \ee^{M_{x,1}(p,\rho_2-\rho_1)}-1 \right)\right\|_1\nonumber \\
&\le& \left\|\Ac\{\rho_1(p,x)\}\left|  \frac{K_2(p)}{K_1(p)} \ee^{M_{x,1}(p,\rho_2-\rho_1) }-1 \right|\right\|_1\nonumber \\
&\le& \|\Ac\{\rho_1(p,x)\}\|_1 \sup_{(p,x)} \left|  \frac{K_2(p)}{K_1(p)} \ee^{M_{x,1}(p,\rho_2-\rho_1) }-1 \right|\nonumber \\
&=& \sup_{(p,x)} \left|  \frac{K_2(p)}{K_1(p)} \ee^{M_{x,1}(p,\rho_2-\rho_1)}-1 \right|
\end{eqnarray}
Now, observe that
\begin{eqnarray}
|M_{x,1}(p,\rho_2-\rho_1)|
&=&  \left|\frac{2}{\sigma^2}\int_0^{x} \mu_{x,1}(p,y, \rho_2-\rho_1)  \dd y\right| \nonumber \\
&\le&  \frac{2}{\sigma^2}\left|\int_0^{x} |\mu_{x,1}(p,y, \rho_2-\rho_1)|  \dd y\right|\nonumber
\end{eqnarray}
and
\begin{eqnarray}
&&\hspace{-4ex} |\mu_{x,1}(p,x, \rho_2-\rho_1)| \nonumber \\ 
&=& \left| \bar{\alpha}(p) \int_{\Zc} \zeta(|x'-x|,p,p') (x'-x) R(p',x')\dd x' \dd p' \right|\nonumber \\
&\le&  \bar{\alpha}(p) \int_{\Zc}\zeta(|x'-x|,p,p') |x'-x| \left|R(p',x')\right|
\dd x'\dd p' \nonumber \\ 
&\le &  S_\zeta  X_0 \|R(p,x)\|_1 \nonumber \\
\end{eqnarray}
where $R(p,x) = \rho_2(p,x)-\rho_1(p,x)$, we defined 
\[S_\zeta  = \sup_{x,x'\in \Xc} \zeta(|x'-x|,p,p')<\infty\]
and  assumed $\zeta(p,p',|x-x'|)=0$ whenever $|x-x'|>X_0$.
Moreover, $\sup\{1-\alpha(p)\}\le 1$ since $0\le\alpha(p)\le 1$.
We also observe that by definition $\|\rho(p,x)\|_1=1$ for any distribution $\rho(p,x)$. Thus $\forall x\in \Xc$
\begin{eqnarray}
|M_{x,1}(p,\rho_2-\rho_1)| 
&\le& \frac{2}{\sigma^2}\left| \int_0^x 2\bar{\alpha}(p)  S_\zeta   X_0 \|R(p,x)\|_1\dd y\right| \nonumber \\
&= & \frac{2}{\sigma^2}  S_\zeta X_0  \|R(p,x)\|_1\left| \int_0^x\dd y\right| \nonumber \\
&\le & \frac{2}{\sigma^2}  S_\zeta X_0 |x|  \|R(p,x)\|_1\nonumber \\
&\le & \frac{2}{\sigma^2}  S_\zeta X_0 S_x   \|R(p,x)\|_1\nonumber \\
&=& \Delta  \label{eq:Delta}
\end{eqnarray}
where $S_x = \sup_{x\in\Xc} |x| <\infty$.  The above inequality
implies that uniformly on $x$ we have:
$-\Delta \le M_{x,1}(p,\rho_2-\rho_1) \le \Delta$, i.e.,
\[  \ee^{-\Delta} \le \ee^{M_{x,1}(p,\rho_2-\rho_1)} \le \ee^{\Delta} \,.\]
Furthermore, observe that from~\eqref{eq:K2} we can obtain the following bounds on
the normalization factor $K_2(p)$:
\begin{eqnarray}
K_2(p)^{-1} 
&=& \int_\Xc \ee^{M_{x,1}(p, \rho_1)}\ee^{-\frac{\alpha(p) [x-u(p)]^2}{\sigma^2}} \ee^{M_{x,1}(p,\rho_2-\rho_1)} \dd x \nonumber \\
&\le& K_1(p)^{-1} \ee^\Delta
\end{eqnarray}
and
\begin{eqnarray}
K_2(p)^{-1} 
&=& \int_\Xc \ee^{M_{x,1}(p, \rho_1)} \ee^{-\frac{\alpha(p) [x-u(p)]^2}{\sigma^2}} \ee^{M_{x,1}(p,\rho_2-\rho_1)} \dd x \nonumber \\
&\ge& K_1(p)^{-1} \ee^{-\Delta}
\end{eqnarray}
which can be summarized as
\[ \ee^{-\Delta} \le \frac{K_2(p)}{K_1(p)} \le \ee^{\Delta} \,. \]
Note also that, for any $|w|<1/2$, we have $\ee^{2|w|}-1\le 4|w|$ and $1-\ee^{-2|w|} \le 2|w|$.
It follows that for $\Delta<1/2$ the term $|K_2(p)/K_1(p)\ee^{M_{x,1}(p,\rho_2-\rho_1)}-1|$ can be bounded as
\begin{eqnarray}
  \left|  \frac{K_2(p)}{K_1(p)} \ee^{M_{x,1}(p,\rho_2-\rho_1)}-1 \right|
&\le& \max\left\{ \ee^{2\Delta}-1,1-\ee^{-2\Delta} \right\} \nonumber \\
&\le& \max\left\{ 4\Delta,2\Delta \right\} \nonumber \\
&\le& 4\Delta \nonumber \\
&=& \frac{8}{\sigma^2} S_\zeta  S_x X_0\|R(p,x)\|_1 \,.
\end{eqnarray}
In conclusion, we get
\begin{equation} \label{eq:g-contraction}
\|\Ac\{\rho_2(p,x)\}-\Ac\{\rho_1(p,x)\}\|_1 \le  \frac{8}{\sigma^2} S_\zeta  S_x X_0\|R(p,x)\|_1 
\end{equation}
Therefore, $\Ac\{\cdot\}$ is a contraction operator if   
\[ \frac{8  S_\zeta    S_x X_0}{\sigma^2} < 1 \]
which implies $\frac{ S_\zeta    S_x X_0}{\sigma^2} < \frac{1}{8}$.
Note that, by the definition of $\Delta$ given in~\eqref{eq:Delta}, we have
\begin{eqnarray}
\Delta 
&=& \frac{2 S_\zeta    S_x X_0}{\sigma^2}\|R(p,x)\|_1\nonumber \\
&\le& 2\frac{1}{8}\|R(p,x)\|_1 \nonumber \\
&\le& \frac{1}{2}
\end{eqnarray}
since $\|R(p,x)||_1 \leq \|\rho_2(p,x)\|_1+\|\rho_1(p,x)||_1\le 2$.
Thus, the condition $\Delta<\frac{1}{2}$ is met.

Following the same approach, we can easily prove that $\Ac\{\cdot\}$ is a
contraction locally in a neighborhood of $\rho^*(p,x)$ (i.e., the fixed point
of $\Ac\{\cdot\}$) whenever $\frac{2   S_\zeta 
  S_x}{\sigma^2} <1- \epsilon \quad \forall \epsilon>0$. {To this end, it is
enough to note that for sufficiently small $w$ and $\forall \epsilon>0$, $e^w <
1+ (1+\epsilon) w$ and that $w/(1-w)< (1+ \epsilon)w$.

At last, observe that from
\eqref{eq:g-contraction} we can immediately deduce  the continuity at every point
of operator $\Ac\{\cdot\}$ w.r.t. the  convergence in norm $\mathbb{L}_1$, provided that $S_\zeta<\infty$ and    $S_x<\infty$ 
(and noticing that by construction  $X_0\le 2S_x$).  
Indeed, as $\|\rho_2(p,x)-\rho_1(p,x)\|_1\to 0$, necessarily $\|\Ac\{\rho_2(p,x)\}-\Ac\{\rho_1(p,x)\}\|_1\to 0$.


\section{Derivation of \protect\eqref{eq:solution}}\label{app:FPsolution}
Consider the FP equation in~\cite[eq. (2)]{OurPaper}, which, thanks to~\cite[eq. (24)]{OurPaper}, can be written as 
\begin{equation} 
\begin{split}
\frac{\partial }{\partial t} \rho(p, x, t)=& \frac{\partial }{\partial x} [(w(p) x - \psi(p,t)) \rho(p, x, t)] + \\
 &\,\,\,\,\,\frac{1}{2}\sigma^2\frac{\partial^2 }{\partial x^2}  \rho(p, x, t)
\end{split}
\end{equation}
where $w(p) = \alpha(p) +\bar{\alpha}(p) \eta(p)$ and $\psi(p,t) =  \bar{\alpha}(p) \eta(p) \phi(p, t)  + \alpha(p) u(p)$. In this appendix, we will omit the argument $p$ unless necessary.

In the following, we will derive the Green function (i.e., the impulse response) of this FP equation for an initial condition $\rho(p, x, 0) = \delta(x-x') \delta(p-p')$.
To this purpose, let us Fourier-transform from variable $x$ to variable $\nu$, obtaining the first-order PDE
\begin{equation} \label{eq:pde_nu}
\left[ \frac{\partial }{\partial t} \mathord{+} w \nu \frac{\partial }{\partial \nu} \right] \widehat{\rho}(p, \nu, t) =
\mathord{-}\left[j 2\pi \nu \psi(t) \mathord{+} 2 \pi^2 \nu^2 \sigma^2 \right] \widehat{\rho}(p, \nu, t).
\end{equation}
Next, we introduce an auxiliary parameter $u$ and consider $t=t(u)$, $\nu=\nu(u)$ and $\widehat{\rho} = \widehat{\rho}(u)$, with initial conditions $t(0) = 0$, $\nu(0) = \nu_0$ and
\begin{equation}
\widehat{\rho}(0) = \widehat{\rho}(p,\nu_0, 0) = e^{-j2\pi \nu_0 x'} 
\delta(p-p')
\end{equation} 
Parameterizing over $u$, we transform the PDE in \eqref{eq:pde_nu} into a system of ODEs by exploiting the identity
\begin{equation} \label{eq:diff_tot}
\left[ \frac{\dd t}{\dd u} \frac{\partial  }{\partial t} + \frac{\dd
    \nu}{\dd u} \frac{\partial  }{\partial \nu} \right]
\widehat{\rho}= \frac{\dd \widehat{\rho}}{\dd u} \,.
\end{equation}
By comparing~\eqref{eq:pde_nu} and~\eqref{eq:diff_tot}, we then get 
\begin{equation}
\left\{
\begin{array}{ccl}
\frac{\dd t}{\dd u} & =&  1 \\
\frac{\dd \nu}{\dd u} & = & w \nu \\
\frac{\dd \widehat{\rho}}{\dd u} & = & -\left[j 2\pi \nu \psi(u) + 2 \pi^2 \nu^2 \sigma^2 \right] \widehat{\rho}
\end{array}
\right.
\end{equation}
which is easily solved as
\begin{equation} \label{eq:t_u}
t(u)  =  u, \,\,\,\, \nu(u)  =  \nu_0 \ee^{w u}
\end{equation}
and
\begin{equation} \label{eq:ode_rhohat}
\frac{\dd \widehat{\rho}}{\dd u} = -\left[j 2\pi \nu_0 e^{w u} \psi(u)
  + 2 \pi^2 \sigma^2 \nu_0^2 e^{2w u}\right] \widehat{\rho}\,.
\end{equation}
The solution of \eqref{eq:ode_rhohat} is
\begin{eqnarray}
\widehat{\rho}(u)  & =& 
\widehat{\rho}(0) \ee^{-j 2\pi \nu_0  \int_0^{u} \ee^{w v}
  \phi(v) \dd v} \exp\left(\pi^2 \sigma^2 \nu_0^2 \frac{1-\ee^{2w u}}{w}\right)  
\nonumber \\
& =& \delta(p-p') \exp\left\{ -j 2\pi \nu_0 \left[x' +
    \int_0^{u} \ee^{w v} \psi(v) \dd v\right] \right . \nonumber \\
&& \left. - \pi^2 \sigma^2 \nu_0^2 \frac{\ee^{2w u}-1}{w}\right\} \,.
 \label{eq:rho_u} 
\end{eqnarray}

By substituting equations~\eqref{eq:t_u} into
\eqref{eq:rho_u}, we finally obtain
\begin{eqnarray*}
  \widehat{\rho}(p,\nu,t|p',x') &=& \delta(p-p')  \exp\left\{-j 2\pi \nu \ee^{-w t} \left[x' \right . \right . \nonumber
\\
&& \hspace{-1cm} \left . \left . +  \int_0^{t} \ee^{w \tau}
    \psi(\tau) \dd\tau \right]
- \pi^2 \sigma^2 \nu^2 \frac{1-\ee^{-2w t}}{w}\right\} \,.
\end{eqnarray*}
Taking the inverse Fourier transform, we get the Green function of the
FP equation as
\begin{eqnarray*}
\rho(p,x,t|p',x') &=& \delta(p-p') \sqrt{\frac{w}{\pi
    \sigma^2 (1-\ee^{-2w t})}} \nonumber \\
&& \hspace{-2cm} \exp\left\{ -\frac{w \left(x -  \ee^{-w t} \left[x' +
   \int_0^{t} \ee^{w \tau} \psi(p,\tau) \dd\tau \right]\right)^2}{\sigma^2 (1-\ee^{-2wt})}\right\}
\end{eqnarray*}
which, as a function of $x$, is a Gaussian pdf with variance
\begin{equation}
\sigma^2(p,t) \triangleq \frac{\sigma^2 (1-\ee^{-2w(p) t})}{2w(p)}
\end{equation}
and mean
\begin{eqnarray*}
m(p,x',t) &\triangleq& \ee^{-w(p) t} \left[x' +  \int_0^{t} \ee^{w(p) \tau} \psi(p,\tau) \dd\tau \right]  \\
&=& e^{-w(p) t} x' + (1-e^{-w(p) t}) \frac{\alpha(p) u(p)}{w(p)}
\\ \nonumber 
&& + \bar{\alpha}(p) \eta(p) \int_0^{t} e^{w(p) (\tau-t)} \phi(p,\tau) d\tau \,.
\end{eqnarray*}
For a general initial density value $\rho(p, x, 0) = \rho_0(x|p)
\rho_0(p)$, we obtain the solution of the FP equation $\rho(p, x,t)$
as 
\begin{equation}
\rho(p,x,t) = \rho_0(p) \int_\Zc \rho(p,x,t|p',x') \rho_0(x'|p') \dd x' \dd p'.
\label{eq:solution}
\end{equation}


\end{document}